\pdfoutput=1

\documentclass[iop]{emulateapj}

\usepackage{natbib,epsfig,siunitx,url,graphicx,amsmath,footnote,float,xcolor,cases}
\usepackage[colorlinks=true,linkcolor=blue,citecolor=blue]{hyperref}

\begin{document}

\submitted{Icarus; accepted}

\title{The early instability scenario: terrestrial planet formation during the giant planet instability, and the effect of collisional fragmentation}

\author{Matthew S. Clement\altaffilmark{1,*}, Nathan A. Kaib\altaffilmark{1}, Sean N. Raymond\altaffilmark{2}, John E. Chambers\altaffilmark{3}, \& Kevin J. Walsh\altaffilmark{4}}

\altaffiltext{1}{HL Dodge Department of Physics Astronomy, University of Oklahoma, Norman, OK 73019, USA}
\altaffiltext{2}{Laboratoire d’Astrophysique de Bordeaux, Univ. Bordeaux, CNRS, B18N, allée Geoffroy Saint-Hilaire, 33615 Pessac, France}
\altaffiltext{3}{Department of Terrestrial Magnetism, Carnegie Institution
for Science, 5241 Broad Branch Road, NW, Washington, DC
20015, USA}
\altaffiltext{4}{Southwest Research Institute, 1050 Walnut St. Suite 300, Boulder, CO 80302, USA}

\altaffiltext{*}{corresponding author email: matt.clement@ou.edu}

\setcounter{footnote}{0}
\begin{abstract}

The solar system's dynamical state can be explained by an orbital instability among the giant planets. A recent model has proposed that the giant planet instability happened during terrestrial planet formation.  This scenario has been shown to match the inner solar system by stunting Mars' growth and preventing planet formation in the asteroid belt. Here we present a large sample of new simulations of the ``Early Instability'' scenario.  We use an N-body integration scheme that accounts for collisional fragmentation, and also perform a large set of control simulations that do not include an early giant planet instability.  Since the total particle number decreases slower when collisional fragmentation is accounted for, the growing planets' orbits are damped more strongly via dynamical friction and encounters with small bodies that dissipate angular momentum (eg: hit-and-run impacts). Compared with simulations without collisional fragmentation, our fully evolved systems provide better matches to the solar system's terrestrial planets in terms of their compact mass distribution and dynamically cold orbits. Collisional processes also tend to lengthen the dynamical accretion timescales of Earth analogs, and shorten those of Mars analogs. This yields systems with relative growth timescales more consistent with those inferred from isotopic dating. Accounting for fragmentation is thus supremely important for any successful evolutionary model of the inner solar system.
\end{abstract}

\section{Introduction}

The ``Nice Model'' \citep{gomes05,Tsi05,mor05} is an evolutionary model for the outer planets that seems to explain many of the solar system's peculiar traits \citep{nesvory07,levison08,nesvorny13,nesvorny15a,nesvorny15b}.  Observations of proto-stellar disks indicate that free gas disappears in a just a few Myr \citep{haisch01,petit01,mamajek08,halliday08,pasucci09}, much faster than the timescales of terrestrial accretion inferred from isotopic dating \citep{currie09,kleine09,Dauphas11}.  Because their gaseous envelopes imply formation in the presence of gas, the outer planets must have formed first \citep{wetherill96,chambers_cassen02,levison_agnor03,raymond04}.  The combined gravitational torques generated by the star, disk and other planets have been shown to quickly force the giant planets into a mutual resonant configuration \citep{masset01,morbidelli07,nesvorny12}.  Such a scenario seems to explain the number of resonant giant exoplanets discovered (eg: Kepler 9, GJ 876 and HR 8799, among others; \citet{holman10,laskar15,Trifonov17,boisvert18}).  After the disk phase of evolution, the scattering of small objects by the outer planets causes the resonant chain to break, and moves the giant planets toward their present orbits.  Saturn, Uranus and Neptune tend to scatter small bodies from the primordial Kuiper Belt inward, while Jupiter preferentially ejects objects from the system \citep{fer84,malhotra93,malhotra95,thommes99}.  To conserve angular momentum, the outermost giant planets' orbits diverge from Jupiter.  When this stable resonant chain eventually collapses, a global instability ensues \citep{gomes05}.  Given that Uranus and Neptune are often ejected in simulations, the current version of the Nice Model invokes the formation of 1-2 additional ice giants in the outer solar system \citep{nesvorny11,nesvorny12}.

The precise timing of the instability is dynamically arbitrary.  Changing the initial conditions of the primordial Kuiper Belt can delay the event.  However, the net result is the same.  A late instability \citep{Tsi05,gomes05,mor05,levison11} would imply a correlation with the late heavy bombardment (LHB, $\sim$700 Myr after gas disk dispersal), the existence of which is now in doubt (for a detailed discussion consult \citet{clement18} and \citet{morb18}).  Moreover, \citet{walshmorb11} favored a late instability when considering constraints for the dynamical structure of the asteroid belt.  On the other hand, an early instability (occurring just a few Myr after the disappearance of the primordial gas disk) is more consistent with Jupiter's binary trojan population \citep{nesvorny18} and the growing ice giant's sculpting of the primordial Kuiper Belt (Ribeiro et al., in prep).  Most importantly, the survival of the terrestrial planets is a very low probability event in a late Nice Model instability \citep{bras09,bras13,agnorlin12,kaibcham16,roig16}.  In \citet{clement18}, henceforward Paper 1, we found that the inner solar system is best reproduced when the instability happens in situ with the process of terrestrial planet formation (roughly 1-10 Myr after the disappearance of the primordial gas disk).  This scenario offers the added benefit of significantly limiting the ultimate mass and formation time of Mars.

Early numerical studies of the late stages (giant impact phase) of terrestrial planet formation successfully reproduced the general orbital spacing of the inner planets \citep{chambers98,chambers01}.  However, these simulations fell short in replicating the low masses of Mercury and Mars (5$\%$ and 10$\%$ that of the Earth, respectively; \citet{wetherill91}), leaving behind a low mass asteroid belt (often Mars to Earth-massed planets were formed in the belt region) and reproducing the dynamically cold orbits of the actual terrestrial system (all inner planets but Mercury have e$\lesssim$0.1 and i$\lesssim 2^{\circ}$).  \citet{obrien06} and \citet{ray06} showed that accounting for the dynamical friction of small planetesimals helped keep the final planet's eccentricities and inclinations low.  However, under-excited systems similar to the solar system still represent a low-likelihood event in most all studies of terrestrial planet formation.  For example, systems in Paper 1 only satisfied this constraint 5-20$\%$ of the time, depending on the initial conditions.  

Potential solutions to the small Mars problem are numerous within the literature (for recent reviews consult \citet{morbray16} and \citet{ray18_rev}).  In Paper 1, we provided a thorough discussion of the major competing models, and argued that any complete model for the evolution of the terrestrial planets must reconcile their survivability within the violent Nice Model instability.  Here, we summarize 3 leading ideas:

$\textbf{1. The ``Grand Tack'' hypothesis:}$ \citet{hansen09} noted that the low masses of Mercury and Mars could be consistently replicated if the initial terrestrial forming disk is confined to a narrow annulus between $\sim$0.7 - 1.0 au.  The ``Grand Tack'' hypothesis \citep{walsh11,walsh16} provides the physical motivation for these initial conditions by surmising that, during the gas phase of evolution, Jupiter migrated in to, and subsequently back out of the inner solar system.  When Jupiter reverses direction (or ``tacks'') at the correct location, the terrestrial disk is truncated at $\sim$ 1.0 au.  However, the mechanism for Jupiter's tack is dependent on the unknown disk structure and gas accretion rates \citep{raymorb14}.

$\textbf{2. An early instability:}$  \citet{ray09a} first recognized that a small Mars could be formed if Jupiter and Saturn were placed in an initial configuration with high eccentricities ($e_{J}$ = $e_{S}$ = 0.1) and mutual inclination ($1.5^{\circ}$).  However, since planet-disk interactions damp out the orbits of the growing giant planets \citep{papaloizou00,tanaka04}, these initial conditions seemed unlikely.  Nevertheless, the giant planet's influence was the crux of many subsequent studies of how Mars' mass can be limited (eg: resonance sweeping \citep{thommes08,bromley17} and resonance crossing \citep{lykawaka13}).  In the early instability scenario of Paper 1, the onset of the Nice Model within a few Myr of gas disk dispersion (as opposed to $\sim$700 Myr in a late version of the Nice Model; \citet{gomes05}) effectively set Mars' geological formation timescale.  Since Mars' accretion is thought to have been mostly complete within 1-10 Myr \citep{mars,Dauphas11}, and Earth's growth continued for $\sim$50-150 Myr \citep{earth,kleine09}, an early instability provides a natural explanation for the disparity in growth timescales.

$\textbf{3. Low mass asteroid belt:}$ 
\citet{iz14} demonstrated that the terrestrial system's orbits could be generated from a steep initial radial mass distribution.  \citet{ray17} expanded on the model \citep{izidoro15,ray17sci} by showing that an empty asteroid belt could be populated with volatile-rich material via aerodynamic drag destabilization of planetesimals during Jupiter's growth phase.  Furthermore, these initial conditions, wherein the asteroid belt and Mars-forming region never contained much material in the first place, are largely consistent with modern pebble accretion simulations \citep{levison15_gp,levison15,draz16}.  However, it is still unclear whether such a steep radial distribution of solids is realistic.

Thus the asteroid belt's low mass is somewhat entangled in the small Mars problem as it is another piece of the inner solar system that numerical simulations struggle to replicate.  As with the small Mars problem, many authors have offered explanations for the asteroid belt's dynamical state \citep{petit01,obrien07,walshmorb11,deienno16,ray17sci,ray17,deienno18}.  In a study similar to Paper 1, \citet{deienno18} analyzed the effects of an early, ``Jumping Jupiter''\footnote[1]{A potential solution to the problem of the terrestrial planets orbits being over-excited in simulations of the Nice Model is forcing a rapid jump in the semi-major axes of Jupiter and Saturn (typically achieved by dynamically scattering an Ice Giant on to a highly-eccentric or hyperbolic orbit).  Thus Jupiter and Saturn ``jump'' across their mutual 2:1 MMR, rather than migrate smoothly through it.  Through this process, the terrestrial system is less disturbed.} style instability on a terrestrial disk composed of 10,000 massless bodies.  The authors found good matches to the actual asteroid belt's structure and composition.  \citet{deienno18} concluded that the early instability scenario proposed by Paper 1 is viable within the constraints of the asteroid belt, however the total depletion in the main belt is insufficient by about two orders of magnitude (thus the authors favor the low primordial massed asteroid belt model of \citet{izidoro15} and \citet{ray17sci}).  The simulations in that paper, however, do not include asteroid self-gravity or the effects of collisional fragmentation.  \citet{clement18_ab} investigated the instability's effect on 3,000, fully self-gravitating asteroids and reported depletions of order 99-99.9$\%$ when the giant planets final orbits most closely matched their current configuration.  Therefore, a primordially depleted asteroid belt might not be necessary within the early instability framework of Paper 1.  The simulations presented in the subsequent sections of our manuscript lack the particle resolution to study the asteroid belt in sufficient detail.  In section 3.3, however, we comment on the effects of collisional fragmentation in the asteroid belt as compared with the results of Paper 1.

A significant limitation of Paper 1 (and most other numerical studies of terrestrial planet formation) was the treatment of all collisions as perfectly accretionary.  \citet{lands12} and \citet{genda12} mapped the various regimes of collisional parameter space, thereby allowing traditional N-body integration packages \citep{duncan98,chambers99} to be modified to provide an approximation for the effects of fragmentation.  \citet{chambers13} conducted the first such study, and found the fully evolved systems of planets to be less dynamically excited than those formed using traditional integration schemes due to angular momentum exchange during hit and run collisions.  Subsequent authors have used similar codes to study various systems \citep{dwyer15,bonsor15,carter15,leinhardt15,quintana15,wallace17}.  However, the sample size of such analyses of planet formation in our solar system remains extremely small.  In this paper, we repeat the simulations of Paper 1 using a version of the $\it{Mercury6}$ hybrid integrator that is modified to handle fragmenting collisions \citep{chambers13}.  In Paper 1 we posited that an early giant planet instability is potentially compatible with other evolutionary schemes that pre-suppose a prior depleted outer terrestrial disk (the Grand Tack and low mass asteroid belt models).  In this paper, we perform an additional suite of simulations where the terrestrial planets form out of a narrow annulus of material (both with and without a Nice Model instability) using the collisional fragmentation scheme.  Thus we provide a side-by-side comparison of i) standard initial conditions that assume no prior depletion in the outer disk, ii) standard initial conditions with fragmentation, iii) standard initial conditions with an instability, iv) standard initial conditions with an instability and fragmentation, v) an annulus with fragmentation and vi) an annulus with fragmentation and an instability. 

\section{Methods}

\subsection{Collisional Algorithm}

The scheme we utilize for approximating the effects of collisional fragmentation \citep{chambers13} is limited by the necessity of setting a minimum fragment mass (MFM).  Setting too low of a MFM will cause the number of particles in the simulation to rapidly multiply, and make the calculation time unreasonable.  Furthermore, collisions producing $\gtrsim$ 90 fragments can overload the Bulirsch-Stoer portion of $\it{Mercury's}$ hybrid-symplectic integrator \citep{wallace17}.  After multiple simulations failed when using MFMs of 0.001 and 0.0025 $M_{\oplus}$, we found a MFM of 0.0055 $M_{\oplus}$ (around half a lunar mass, or D$\approx$2000 km assuming $\rho=$ 3.0 $g/cm^{3}$) to be a good choice for 200 Myr simulations of terrestrial evolution.  Detailed descriptions of the identical fragmentation scheme can be found in \citet{chambers13} and \citet{dwyer15}.  In general, when a collision is detected, the mass of the largest remnant is calculated utilizing relations from \citet{lands12}.  Any leftover mass is assigned to a set of equal-massed fragments, each with masses greater than the MFM.  The fragments are then ejected in random, uniform directions within the collisional plane at $\sim$5$\%$ greater than the two body escape velocity. 

\subsection{Giant Planet Configurations}

\begin{table*}
\centering
\begin{tabular}{c c c c c c c c}
\hline
Name  & $N_{Pln}$ &  $M_{disk}$ & $\delta$r & $r_{out}$ & $a_{nep}$ & Resonance Chain & $M_{ice}$\\
& & ($M_{\oplus}$) & (au) & (au) & (au) & & ($M_{\oplus}$)\\
\hline
5GP & 5 & 35 & 1.5 & 30 & 17.4 & 3:2,3:2,3:2,3:2 & 16,16,16 \\
6GP & 6 & 20 & 1.0 & 30 & 20.6 & 3:2,4:3,3:2,3:2,3:2 & 8,8,16,16 \\
\hline
\end{tabular}
\caption{Table of giant planet initial resonant configurations from Paper 1.  The columns are: (1) the name of the simulation set, (2) the number of giant planets, (3) the mass of the planetesimal disk exterior to the giant planets, (4) the distance between the outermost ice giant and the planetesimal disk’s inner edge, (5) the semi-major axis of the outermost ice giant (commonly referred to as Neptune, however not necessarily the planet which completes the simulation at Neptune's present orbit), (6) the resonant configuration of the giant planets starting with the Jupiter/Saturn resonance, and (7) the masses of the ice giants from inside to outside.}
\label{table:gp}
\end{table*}

We described the evolution of our giant planet resonant chains in detail in Paper 1.  Since the parameter space of possible primordial orbits for the outer planets is exhaustive, we use the most successful 5 and 6 planet configurations from \citet{nesvorny12} (denoted n1 and n2 in Paper 1, henceforward 5GP and 6GP for simplicity; table \ref{table:gp}).  The giant planets are migrated into the appropriate configuration using an additional force designed to approximate gas disk interactions by modifying the equations of motion with forced migration ($\dot{a}$) and eccentricity damping ($\dot{e}$) terms \citep{lee02,clement17}.  The resonant chains are then integrated with a 20 day time-step in the presence of 1000 equal-mass primordial Kuiper Belt objects (see Paper 1) up until the point when two giant planets first pass within 3 mutual Hill Radii.  The terrestrial disks (section 2.3) are then added, and the complete system is integrated through the instability using a 6.0 day time-step.

By its nature, the instability is a chaotic event.  Since the parameter space of possible final giant planet configurations is so extensive, outcomes that replicate the actual solar system in broad strokes are quite rare.  In fact, only around one third of the best simulation sets in \citet{nesvorny12} finish the integration with the correct number of planets.  In Paper 1, we analyzed all the different outer solar system outcomes, and found that the instability's tendency to limit the mass and formation time of Mars is largely independent of the particular outcome of the instability.  However, we noted that the most successful terrestrial outcomes occurred when the evolution of the giant planets was most akin to that of the solar system in terms of the final period ratio of Jupiter and Saturn, and the excitation of Jupiter's $g_{5}$ mode.  For these reasons, we stop simulations if Jupiter and Saturn's period ratio ever exceeds 2.8, or if an ice giant is not ejected within 5 Myr.  In the analysis sections of this manuscript (section 3), we only compare our new fragmentation systems to our Paper 1 systems with Jupiter and Saturn inside a period ratio of 2.8.

\subsection{Terrestrial Disks}

\subsubsection{Standard initial conditions}

We follow the same general approach for testing the instability's effect on terrestrial planet formation as in Paper 1.  We begin by studying the classic model of terrestrial planet formation \citep{chambers01,obrien06,ray06,ray09a}.  This model, which we refer to as our ``standard'' set of initial conditions, assumes that the terrestrial planets formed out of a disk of large, Moon-Mars massed planet-forming embryos, and smaller planetesimals, extending between the present location of Mercury and the asteroid belt's outer edge.  Thus, our runs utilizing standard initial conditions assume no prior depletion in the outer terrestrial disk (2.0-4.0 au).  Our simulations begin by following the collisional evolution of 1000 equal-mass planetesimals and 100 equal-mass planet embryos \citep{obrien06} in the presence of a non-migrating Jupiter and Saturn.  Planetesimals only interact gravitationally and undergo collisions with the embryos.  Fragments produced in collisions between embryos are treated as fully self-gravitating, while fragments generated in planetesimal-embryo collisions only interact with embryos.  Jupiter and Saturn are placed on their pre-instability orbits in a mutual 3:2 mean motion resonance (MMR, $a_{j}$=5.6 au, $a_{s}$=7.6 au), a typical outcome of Jupiter and Saturn's migration during the gas disk phase \citep{masset01,morbidelli07,pierens14}.  The initial terrestrial disk mass is set to 5 $M_{\oplus}$, and divided evenly between the embryos and planetesimals.  The disk boundaries are at 0.5 and 4.0 au, and the spacing between objects is selected to achieve a surface mass density profile proportional to $r^{-3/2}$.  Angular orbital elements are drawn from random, uniform distributions.  Eccentricities and inclinations are selected randomly from near circular, gaussian distributions ($\sigma_{e}=.02$ and $\sigma_{i}=.2^{\circ}$).  These initial conditions are selected for their simplicity, and consistency with previous works \citep{chambers01,ray06,obrien06,ray09a,kaibcowan15}.  In Paper 1 we tested different inner disk edges and total disk masses, and found that disks smaller than 5 $M_{\oplus}$ systematically failed to produce Earth and Venus analogs with the correct masses.  However, the location of the inner disk edge appeared statistically uncorrelated with any of our success criteria.  For these reasons, we do not vary any of these parameters in this work.  

We perform 100 simulations as described above using our fragmentation integration scheme \citep{chambers99,chambers13}, and a 6 day time-step.  We take snapshots of the terrestrial disks at 1.0, 5.0 and 10.0 Myr to input into our giant planet instability simulations (section 2.2).  This allows us to test different instability timings with respect to the evolutionary state of the terrestrial disk.  In the subsequent text, we refer to these sets collectively as our instability set.  The various grouping of simulations testing different instability delay times are referred to as ``batches,'' and individual simulations are called ``runs.''  The most successful outcomes of Paper 1 occurred in the 1.0 and 10.0 Myr simulation batches, hence our choices of output times.  The same systems are also integrated up to 200 Myr without any giant planet evolution, and become our standard initial condition control set (we refer to this set henceforward as the standard set).

\subsubsection{Annulus initial conditions}

Our annulus set of simulations are set up much the same as the standard initial condition set, with two exceptions.  The annulus runs begin with just 400 embryos (planetesimals are not used) distributed between 0.7 to 1.0 au \citep{hansen09}.  For this batch, we only take a snapshot at 10 Myr for input into a Nice Model instability due to the massive computational requirements of this project.  The annulus runs are also integrated up to 200 Myr without giant planet evolution, and become the annulus control set (referred to henceforward as the annulus set).  Otherwise, the integrator, initial giant planet configuration, time-step and orbital element selection method are the same as described above.  We summarize the initial conditions used in our various integrations in table \ref{table:ics2}.

The low mass asteroid belt \citep{iz14,izidoro15,ray17sci} and Grand Tack \citep{walsh11,jacobson14,rubie15,deienno16,brasser16,walsh16} models both assume a truncated terrestrial disk that is largely depleted of material in the primordial asteroid belt and Mars-forming regions \citep{morbray16}.  Our present study is by no means an exhaustive investigation of either the Grand Tack or low mass asteroid belt scenarios.  We begin our annulus simulations with overly-simplified initial conditions.  Furthermore, our study does not model the inward and outward migration phase of the Grand Tack.  Nevertheless, we still present these simulations in section 3.5 for three reasons.  First, to understand whether accounting for fragmentation is a potential barrier to the success of the annulus setup (also addressed for the Grand Tack in \citet{walsh16}).  Second, to ascertain whether the initial conditions are compatible with the early instability framework of Paper 1 (an open-ended question from that work).  And finally, to study the relative accretion timescales of the planets.

\begin{table*}
\centering
\begin{tabular}{c c c c c c c c c c }
\hline
Set & $a_{in}$ (au) & $a_{out}$ (au) & $M_{tot}$ ($M_{\oplus}$) & $N_{emb}$ & $N_{pln}$ & $N_{GP}$ & $t_{instb}$ (Myr) & $N_{sim}$\\
\hline
standard & 0.5 & 4.0 & 5.0 & 1000 & 100 & 2 & N/A & 100 \\
annulus & 0.7 & 1.0 & 2.0 & 400 & 0 & 2 & N/A & 100\\
standard/5GP/1Myr  & 0.5 & 4.0 & 5.0 & 1000 & 100 & 5 & 1 & 31\\
standard/5GP/5Myr  & 0.5 & 4.0 & 5.0 & 1000 & 100 & 5 & 5 & 8\\
standard/5GP/10Myr  & 0.5 & 4.0 & 5.0 & 1000 & 100 & 5 & 10 & 21\\
standard/6GP/1Myr  & 0.5 & 4.0 & 5.0 & 1000 & 100 & 6 & 1 & 13\\
standard/6GP/5Myr  & 0.5 & 4.0 & 5.0 & 1000 & 100 & 6 & 5 & 30\\
standard/6GP/10Myr  & 0.5 & 4.0 & 5.0 & 1000 & 100 & 6 & 10 & 40\\
annulus/5GP/10Myr & 0.7 & 1.0 & 2.0 & 400 & 0 & 5 & 10 & 27\\
annulus/6GP/10Myr & 0.7 & 1.0 & 2.0 & 400 & 0 & 6 & 10 & 46\\
\hline	
\end{tabular}
\caption{Summary of initial conditions for complete sets of terrestrial planet formation simulations.  The columns are (1) the name of the simulation set, (2) the inner edge of the terrestrial forming disk, (3) the disk's outer edge, (4) the total disk mass, (5) the number of equal-mass embryos used and (6) the number of equal-mass planetesimals used, (7) the number of giant planets, (8) the instability timing in Myr, and (9) the total number of integrations comprising the set (Note that each set begins with 100 calculations, however instability runs where Jupiter and Saturn exceed a period ratio of 2.8 or an ice giant is not ejected within 5 Myr are removed).  In the subsequent text, we often refer to the 6 standard/XGP/XMyr (rows 3-8) collectively as our instability sets.  It should also be noted that the ``standard'' set was referred to as the ``control'' set in Paper 1.}
\label{table:ics2}
\end{table*}

\subsection{Success Criteria}

We employ the same success criteria for our fully formed terrestrial disks as in Paper 1 (table \ref{table:crit}).  Because we remove all systems where the Saturn to Jupiter period ratio is greater than 2.8, or if an ice giant is not ejected within 5 Myr, we do not scrutinize our systems against the \citet{nesvorny12} giant planet success criteria as in Paper 1.  The motivation for the success criteria is described in detail in Paper 1.  Here, we provide a brief synopsis for each criterion we utilize.

\begin{table*}
\centering
\begin{tabular}{c c c c c}
\hline
Code & Criterion &  Actual Value & Accepted Value & Justification\\
\hline
A & $a_{Mars}$ & 1.52 au & 1.3-2.0 au & Inside AB \\
A,A1 & $M_{Mars}$ & 0.107 $M_{\oplus}$ & $>0.025,<.3 M_{\oplus}$ & \citep{ray09a} \\
A,A1 & $M_{Venus}$ & 0.815 $M_{\oplus}$ & $>$0.6 $M_{\oplus}$ & Within $\sim25\%$\\
A,A1 & $M_{Earth}$ & 1.0 $M_{\oplus}$ & $>$0.6 $M_{\oplus}$ & Match Venus\\
B & $\tau_{Mars}$ & 1-10 Myr & $<$10 Myr & \\
C & $\tau_{\oplus}$ & 50-150 Myr & $>$50 Myr & \\
D & $M_{AB}$ & $\sim$ 0.0004 $M_{\oplus}$ & No embryos & \citep{chambers01} \\
E & $\nu_{6}$ & $\sim$0.09 & $<$1.0 & \\
F & $WMF_{\oplus}$ & $\sim10^{-3}$ & $>10^{-4}$ & Order of magnitude\\
G & AMD & 0.0018 & $<$0.0036 & \citep{ray09a} \\
\hline
\end{tabular}
\caption{Summary of success criteria for the inner solar system from Paper 1.  The rows are: (1) the semi-major axis of Mars, (2-4) The masses of Mars, Venus and Earth, (5-6) the time for Mars and Earth to accrete $90\%$ of their mass, (7) the final mass of the asteroid belt, (8) the ratio of asteroids above to below the $\nu_{6}$ secular resonance between 2.05-2.8 au, (8) the water mass fraction of Earth, and (9) the angular momentum deficit (AMD) of the inner solar system.}
\label{table:crit}
\end{table*}

\subsubsection{The Structure of the Inner Solar System}

We use metrics from \citet{chambers01} to scrutinize the general orbital structure of the inner solar system.  Criteria A and A1 quantify the terrestrial planetary system's semi-major axis spacing and mass distribution.  Any planets formed in the region between 1.3-2.0 au are considered Mars analogs.  The inner limit of this region is close to Mars' actual pericenter ($\sim$1.38 au) and the outer edge lies at the asteroid belt's inner limit.  Criterion A is satisfied if the Mars analog is smaller than 0.3 $M_{\oplus}$, exterior to Earth and Venus analogs each with masses greater than 0.6 $M_{\oplus}$, and the asteroid belt is devoid of objects more massive than 0.3 $M_{\oplus}$.  Criterion A1 is similar, but it also includes systems that form no Mars, and those that meet the mass distribution requirement but fail the semi-major axis requirements as being successful.  We also scrutinize the mass distribution of each system statistically, using a normalized radial mass concentration statistic (RMC in equation \ref{eqn:sc}, \citet{chambers01}).  A system of planets dominated by a single, massive planet would yield a steep mass concentration function (the expression in parenthesis) and thus have a high RMC.  A system with many smaller planets, and a smoother mass concentration function, would yield a lower RMC.  We normalize all of our RMC values to the solar system statistic for Mercury, Venus, Earth and Mars ($RMC_{SS}$ = 90). 

\begin{equation}
        RMC = MAX\bigg(\frac{\sum_{i}m_{i}} {\sum_{i}m_{i}[\log_{10}(\frac{a}{a_{i}})]^2}\bigg)
        \label{eqn:sc}
\end{equation}

To quantify the orbital excitation of our terrestrial systems, we calculate the angular momentum deficit (AMD, equation \ref{eqn:amd}) for each planetary system.  AMD \citep{laskar97} computes the degree to which a system of orbits differs from one with circular and co-planar orbits.  Because chaotic dynamics can cause the solar system's AMD to naturally evolve by as much as a factor of two over Gyr integrations \citep{laskar97,agnor17}, we only require our systems achieve an AMD less than twice the solar system's modern value of 0.0018 (criterion G).  For both our AMD and RMC calculations, we only consider planets (Mercury-massed objects and larger; $m>0.055$ $M_{\oplus}$) that are not in the asteroid belt ($a<2$ au).  This allows us to best compare our simulated systems with the actual solar system.  It should also be noted that, given our choice of MFM (one tenth that of our planet cut-off mass), fragments are rarely considered in our RMC and AMD calculations.

\begin{equation}
	AMD = \frac{\sum_{i}m_{i}\sqrt{a_{i}}[1 - \sqrt{(1 - e_{i}^2)}\cos{i_{i}}]} {\sum_{i}m_{i}\sqrt{a_{i}}} 
	\label{eqn:amd}
\end{equation}

\subsubsection{Relative Formation Timescales}

As we discussed in Paper 1, the disparity between Mars' inferred rapid formation timescale of just a few million years \citep{mars,Dauphas11,kruijer17} and Earth's (longer by about a factor of 10, \citet{earth,kleine09}) is an important and defining trait of the inner solar system.  However, there is considerable uncertainty in both of these Hf/W dates \citep{Dauphas11}.  Therefore, we only require our Mars analogs accrete 90$\%$ of their final mass within 10 Myr (criterion B), and our Earth's take at least 50 Myr to do the same (criterion C).

\subsubsection{The Asteroid Belt}

In Paper 1 we argued that standard embryo accretion models have insufficient mass and particle resolution to accurately study the asteroid belt.  In particular, we model the belt with 0.025 $M_{\oplus}$, fully self-gravitating embryos and 0.0025 $M_{\oplus}$ planetesimals that only interact gravitationally with the larger objects (each object being tens to hundreds of times more massive than the entire present belt).  If perturbations from the giant planets are the dominant sculpting mechanisms in the asteroid belt, then we could expect each individual asteroid to behave as a test particle.  Meanwhile, if asteroid belt "self-stirring" is a more important process, then our super-massive asteroid bodies likely overestimate the level of self-stirring.  Furthermore, our choice of initial conditions could also lead to unrealistic fragmentation effects since it requires less energy to break apart a small asteroid than a larger one.  These issues are addressed with more detailed simulations in a complimentary study \citep{clement18_ab}.  Nevertheless, for completeness, we include the same criteria for the asteroid belt as in Paper 1 (an asteroid belt completely depleted of planetary embryos, criterion D, and the ratio of asteroids above to below the $\nu_{6}$ secular resonance between 2.05 and 2.8 au less than 1.0, criterion E). 

\subsubsection{Earth's Water Content}

Studies and models for the origin of Earth's water are numerous in the literature.  A complete discussion of this topic is far beyond the scope of this paper (for robust reviews of various ideas see \citet{morb00}, \citet{morb12} and \citet{marty16}).  As in Paper 1, we assume a simple bulk distribution (equation \ref{eqn:wmf}) of water-rich material similar to \citet{ray09a}.  This distribution assumes that the inner solar system and primordial asteroid belt region was populated with water-rich material from the outer solar system during the gas disk phase \citep{ray17}.  Criterion F is satisfied if an Earth analog's (m$>$0.6 $M_{\oplus}$, 0.85$<$a$<$1.3 au) water mass fraction (WMF) is boosted to greater than $10^{-4}$.

\begin{equation}
WMF=
\begin{cases}
10^{-5},\quad r<2au\\
10^{-3},\quad 2au<r<2.5au\\
10\%,\qquad r>2.5au
\end{cases}
\label{eqn:wmf}
\end{equation}

\section{Results and Discussion}

\begin{figure*}
\centering
\includegraphics[width=.49\textwidth]{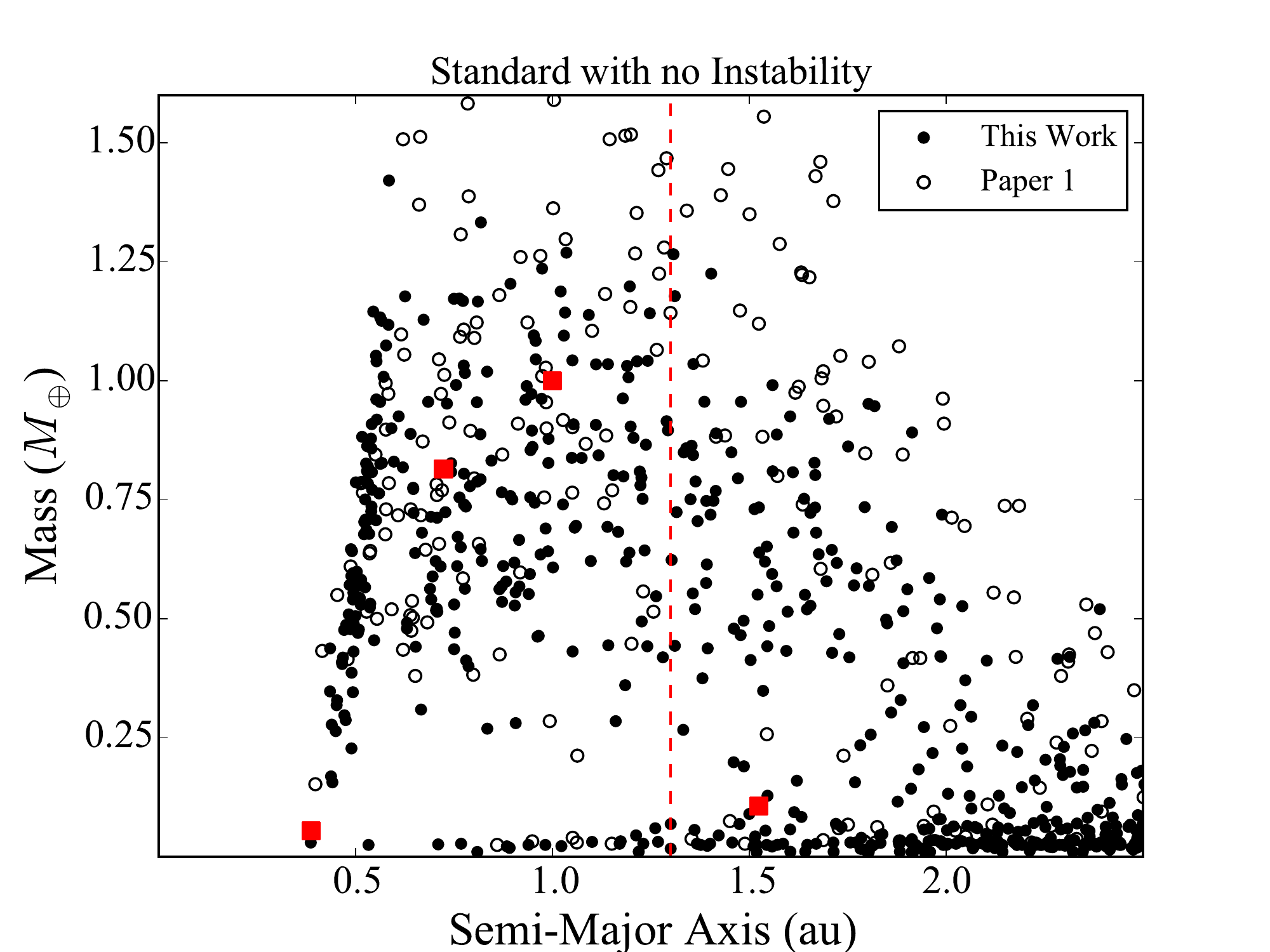}	
\includegraphics[width=.49\textwidth]{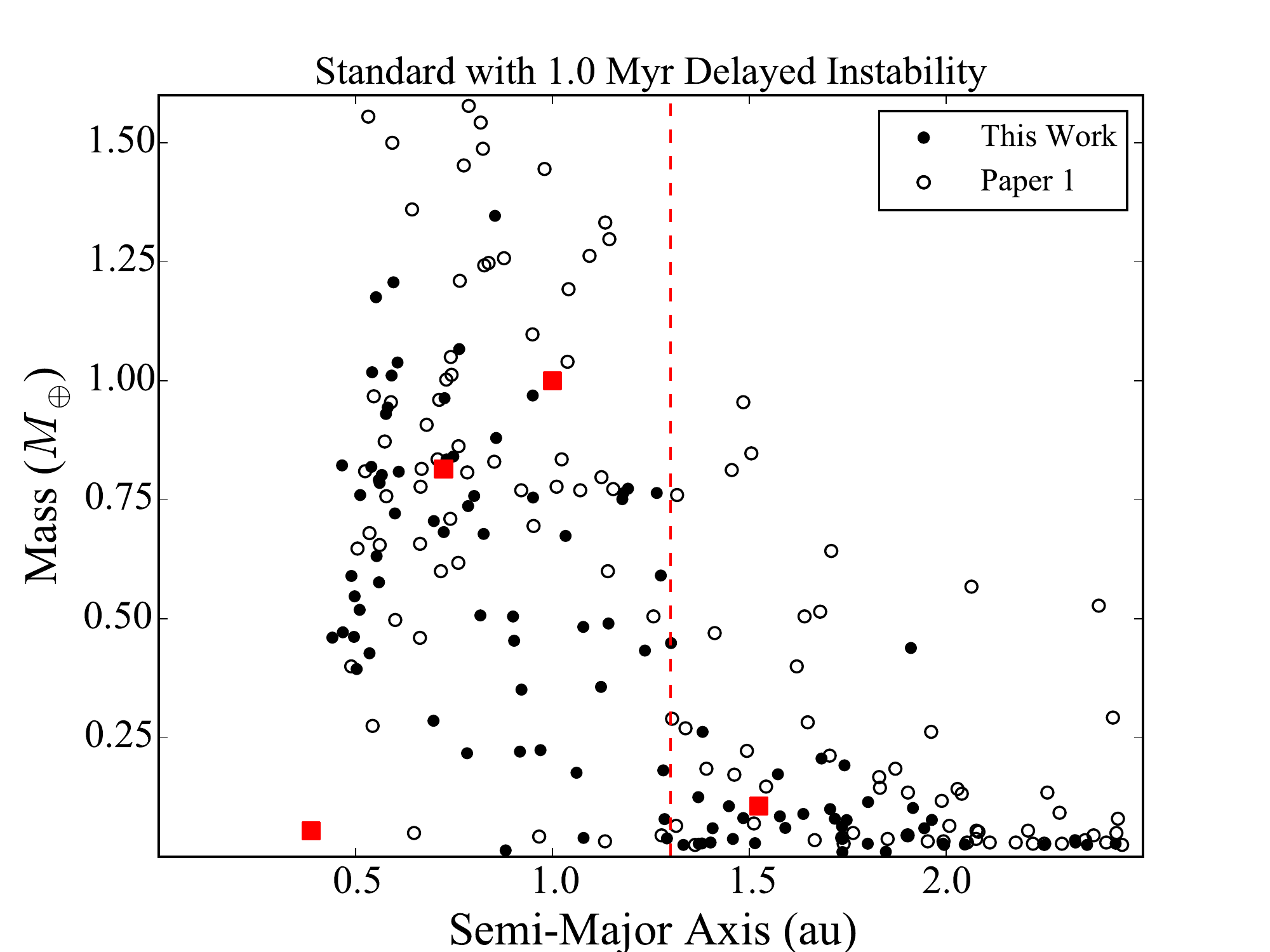}	
\qquad
\includegraphics[width=.49\textwidth]{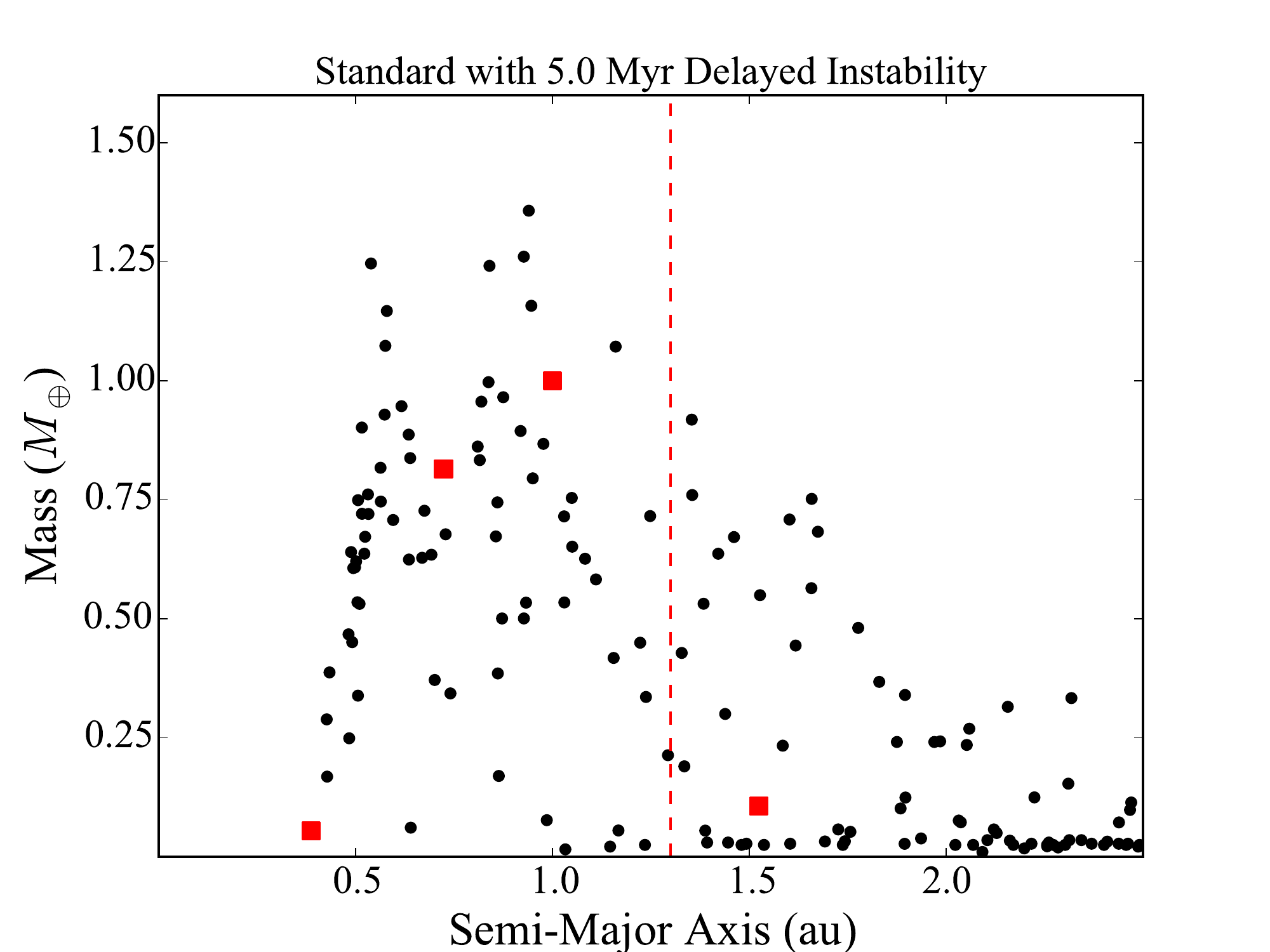}	
\includegraphics[width=.49\textwidth]{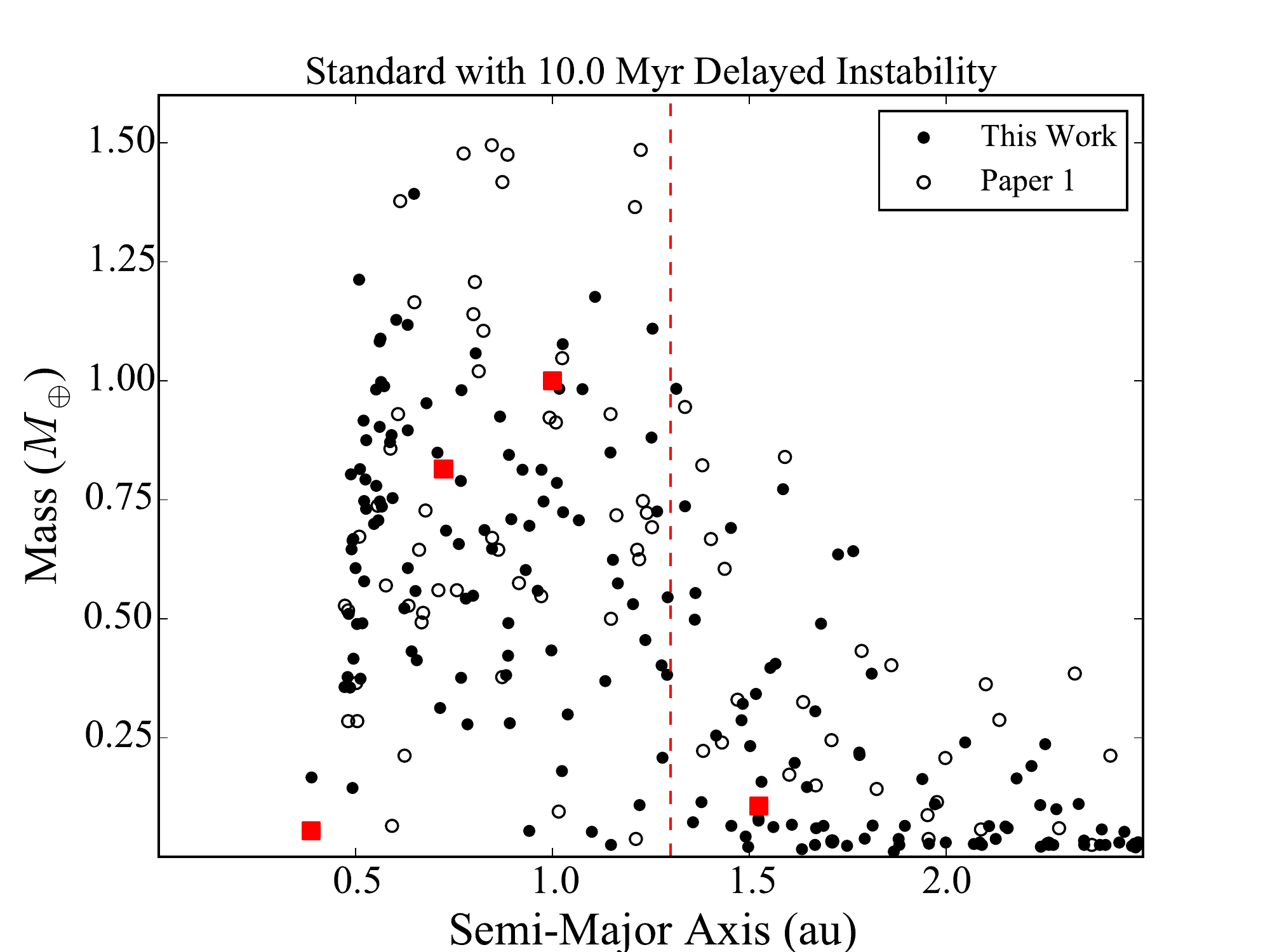}	
\qquad
\includegraphics[width=.49\textwidth]{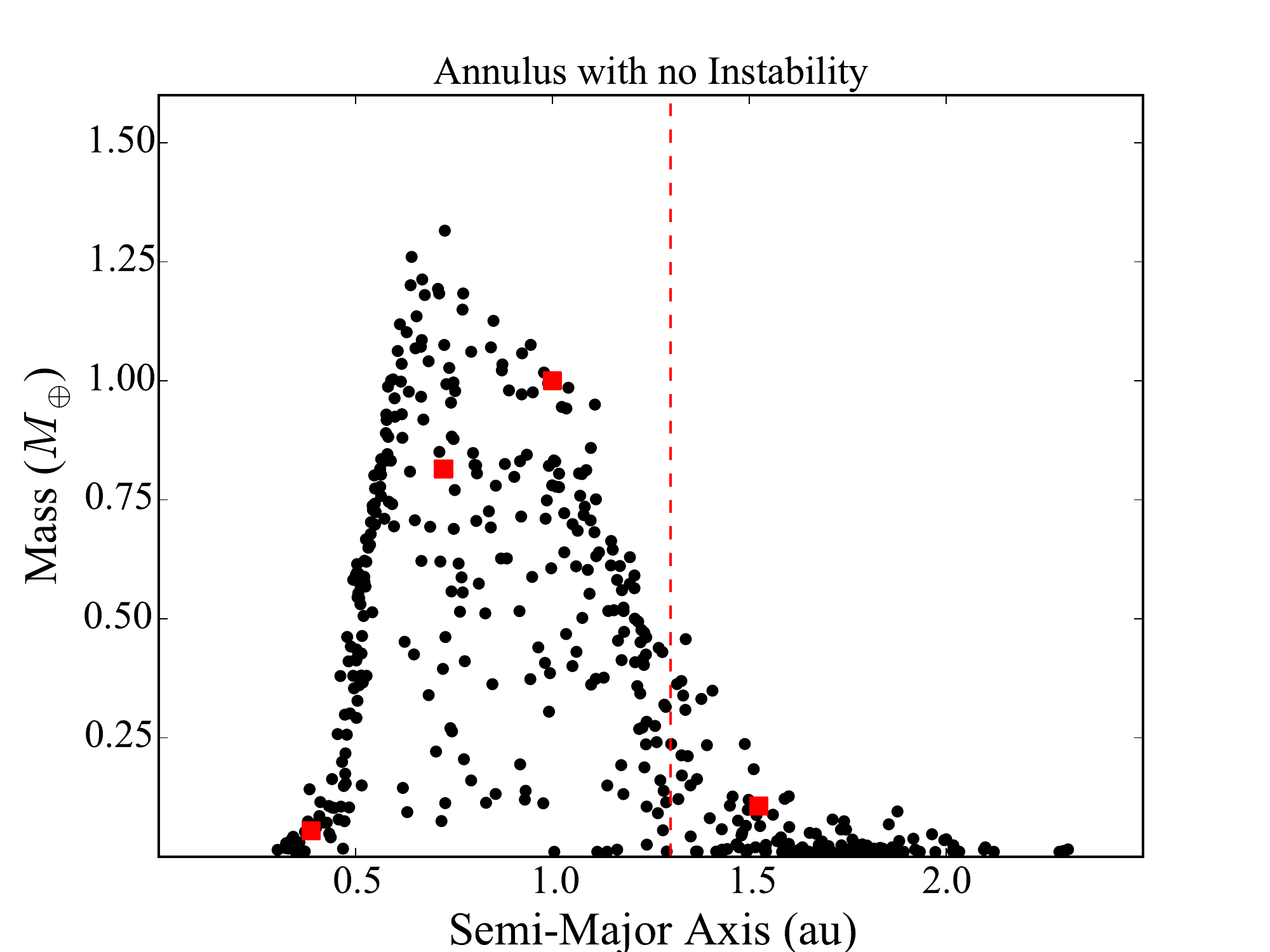}	
\includegraphics[width=.49\textwidth]{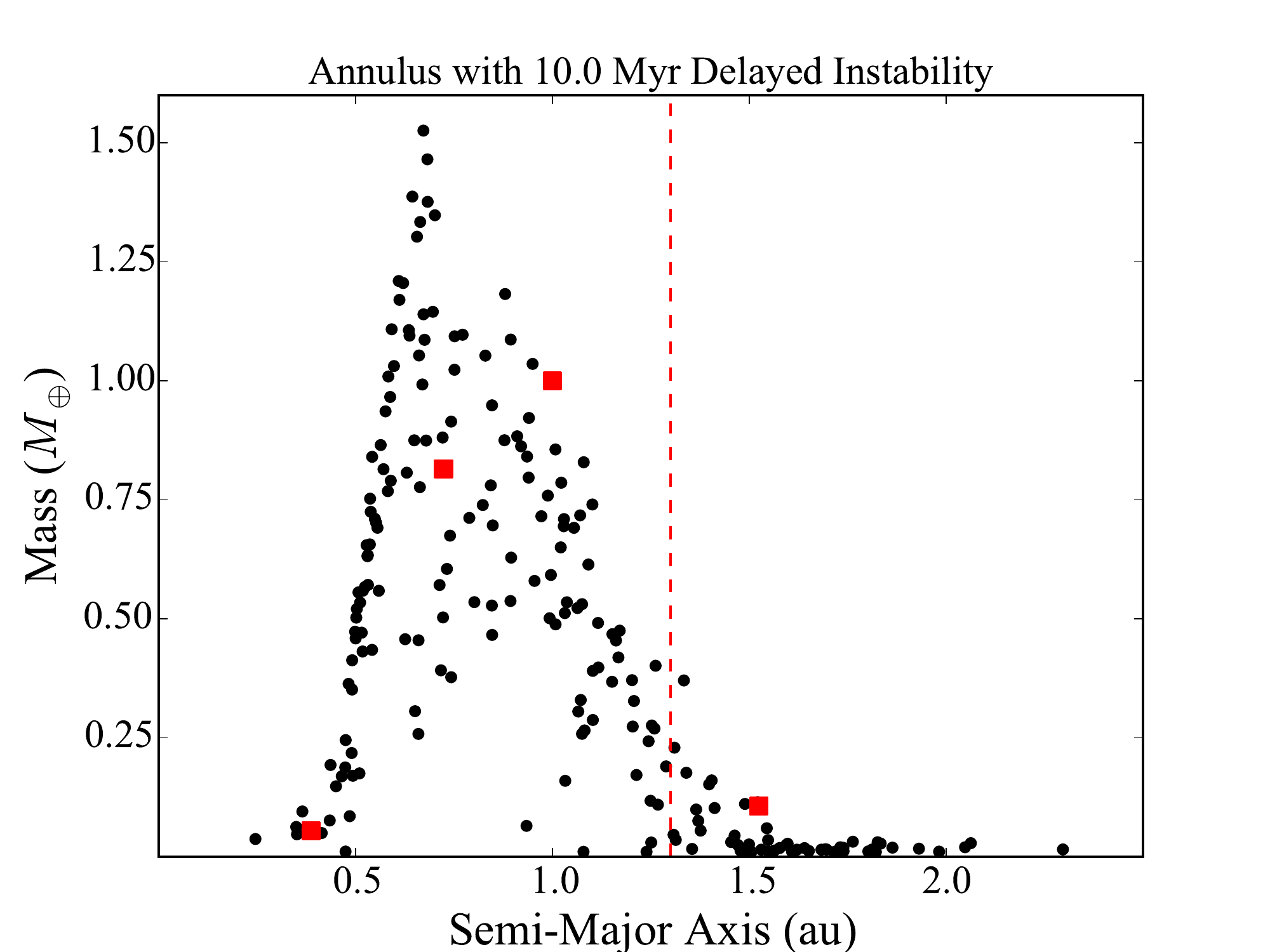}	
\caption{Distribution of semi-major axes and masses for all planets formed in all simulation sets.  The red squares denote the actual solar system values for Mercury, Venus, Earth and Mars.  The vertical dashed line separates the Earth and Venus analogs (left side of the line) and the Mars analogs (right side).  Planets from Paper 1 simulations that do not include collisional fragmentation (control (standard) sets, 1.0 Myr and 10.0 Myr instability delays) are denoted with open circles.}
\label{fig:totalplot}
\end{figure*}

We begin by summarizing the percentages of each simulation subset that meet our various success criteria in a modified version of table 4 from Paper 1 (table \ref{table:results}).  In figure \ref{fig:totalplot}, we plot the distribution of planet masses and semi-major axes for each simulation subset as well.  In this work we stop all simulations where Jupiter and Saturn's period ratio exceeds 2.8.  Additionally, we only test instability delay times of 1.0, 5.0 and 10.0 Myr (as opposed to Paper 1 where we investigated 0.01, 0.1, 1.0 and 10.0 Myr).  Because of these differences, in the subsequent analysis sections we only compare our current results with the Paper 1 systems that tested 1.0 and 10.0 Myr instability delay times, and finished with Jupiter and Saturn within a period ratio of 2.8.

\begin{table*}
\centering
\begin{tabular}{c c c c c c c c c}
\hline
Set & A & A1 & B & C & D & E & F & G\\ 
& $a,m_{TP}$ & $m_{TP}$ & $\tau_{mars}$ & $\tau_{\oplus}$ & $M_{AB}$ & $\nu_{6}$ & WMF & AMD\\
\hline
std & 2 (+2) & 2 (+2) & 33 (+24) & 80 (-5) & 2 (0) & 0 (-53) & 82 (-4) & 15 (+7)\\
ann & 58 & 58 & 51 & 27 & 85 & 3 & N/A & 50\\
std/5GP/1Myr & 35 (+9) & 35 (+9) & 23 (+11) & 80 (-15) & 70 (+50) & 0 (-29) & 40 (-25) & 12 (-1)\\
std/6GP/1Myr & 33 (+22) & 33 (+6) & 25 (+13) & 100 (+8) & 77 (+35) & 11 (-32) & 66 (-2) & 37 (+35)\\
std/5GP/5Myr & 13 & 13 & 50 & 50 & 100 & 13 & 75 & 25\\
std/6GP/5Myr & 17 & 24 & 40 & 100 & 27 & 3 & 91 & 17\\
std/5GP/10Myr & 36 (+27) & 36 (+18) & 28 (+21) & 83 (-8) & 44 (+28) & 4 (-25) & 83 (+30) & 32 (+7)\\
std/6GP/10Myr & 17 (-2) & 17 (-15) & 50 (+23) & 77 (-9) & 47 (+22) & 0 (-52) & 77 (+0) & 29 (+27)\\
ann/5GP/10Myr & 22 & 22 & 64 & 20 & 94 & 0 & N/A & 17\\
ann/6GP/10Myr & 29 & 32 & 76 & 12 & 97 & 2 & N/A & 29\\
\hline
\end{tabular}
\caption{Summary of percentages of systems which meet the various terrestrial planet success criteria established in table \ref{table:crit}.  Values in parenthesis indicate the change from the same simulation set in Paper 1 (where fragmentation was not considered).  The subscripts TP and AB indicate the terrestrial planets and asteroid belt respectively.}
\label{table:results}
\end{table*}

\subsection{The Small Mars Problem}

\begin{figure}
	\centering
	\includegraphics[width=.50\textwidth]{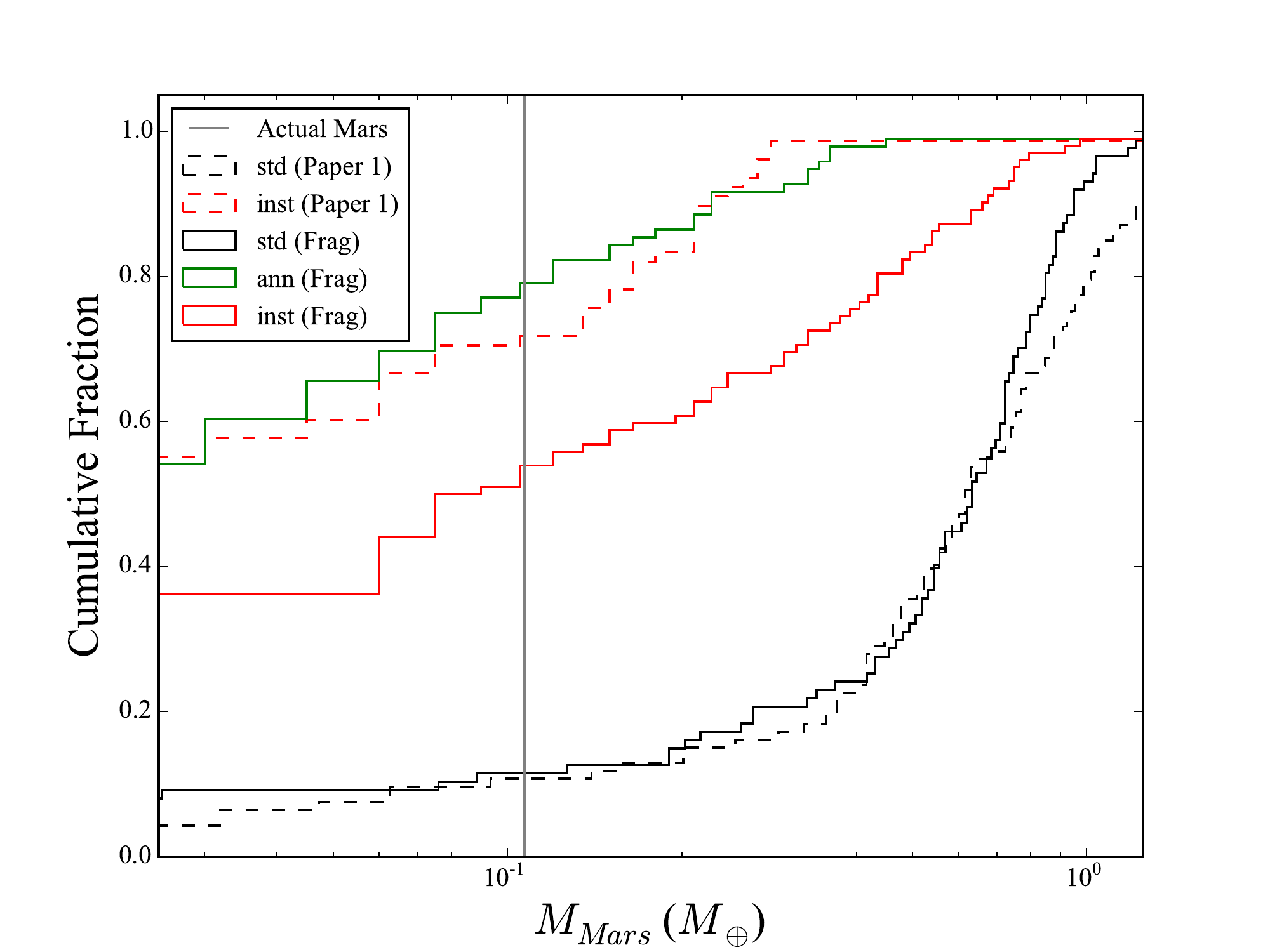}
	\qquad
	\includegraphics[width=.50\textwidth]{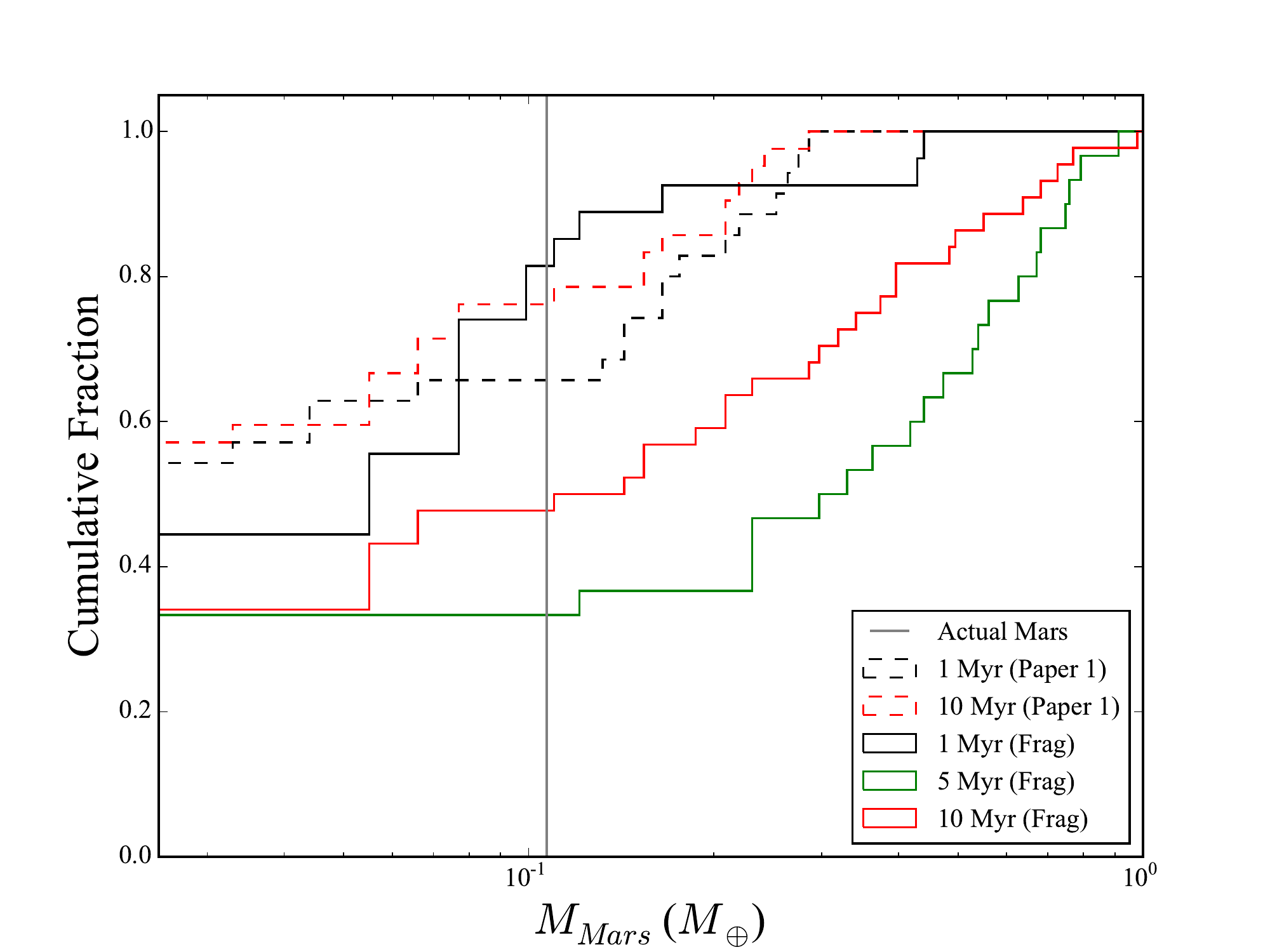}
	\caption{The top panel depicts the cumulative distribution of Mars analog masses formed in our various fragmentation simulation sets (solid lines) compared with our results from Paper 1 (dashed lines). The bottom panel depicts the same data for the various instability delay times tested in this work (solid lines), and in paper 1 (dashed lines).  The grey vertical lines corresponds to Mars’ actual mass.  Note that some systems may form multiple planets in this region, but here we only plot the most massive planet.  Systems that do not form a Mars analog via embryo accretion are plotted as having zero mass.}
	\label{fig:mars}
\end{figure}

Our fragmentation simulations perform better than the systems from Paper 1 when measured against some of our success criteria, but perform worse when scrutinized against others.  As evidenced by criterion A and A1 success rates of $\sim$15-35$\%$ (table \ref{table:results}), an orbital instability still seems to be more efficient at replicating the actual terrestrial system than standard initial conditions without any giant planet evolution.  The most obvious difference between the terrestrial systems formed in this work, and those from Paper 1, is that when fragmentation is included, the simulations produce consistently larger Mars analogs (figure \ref{fig:totalplot}).  Indeed, when we compare the cumulative distributions of Mars analogs formed across our various simulation sets (figure \ref{fig:mars}) we find that the addition of the fragmentation algorithm leads to a marked increase in Mars masses for our instability runs.  This is particularity the case in our batches testing later instability delay times (5.0 and 10.0 Myr).  Though systems in our 1.0 Myr instability delay batches consistently produce the best Mars analogs (figures \ref{fig:totalplot} and \ref{fig:mars}), they are also more likely to destroy Mars all together.

In Paper 1 we argued that earlier instability delay times (0.01 and 0.1 Myr) are less successful at generating a small Mars.  When the instability occurs earlier in the process of terrestrial planet formation, a greater fraction of the total disk mass is distributed in a sea of small planetesmials.  The increased dynamical friction between the planetesimals and growing embryos that remain after the instability truncates the disk has a net spreading effect.  While the instability does remove mass from the Mars-forming region, the area is essentially repopulated with material as the mass distribution profile flattens via dynamical friction and scattering events.  Furthermore, the increased dynamical friction in the vicinity of the growing Mars tends to dampen orbits in the area, and prevents the unstable giant planets from exciting material on to orbits where it is lost from the system.  These combined effects tend to yield over-massed Mars and under-massed Earth and Venus analogs.

\begin{figure}
	\centering
	\includegraphics[width=.50\textwidth]{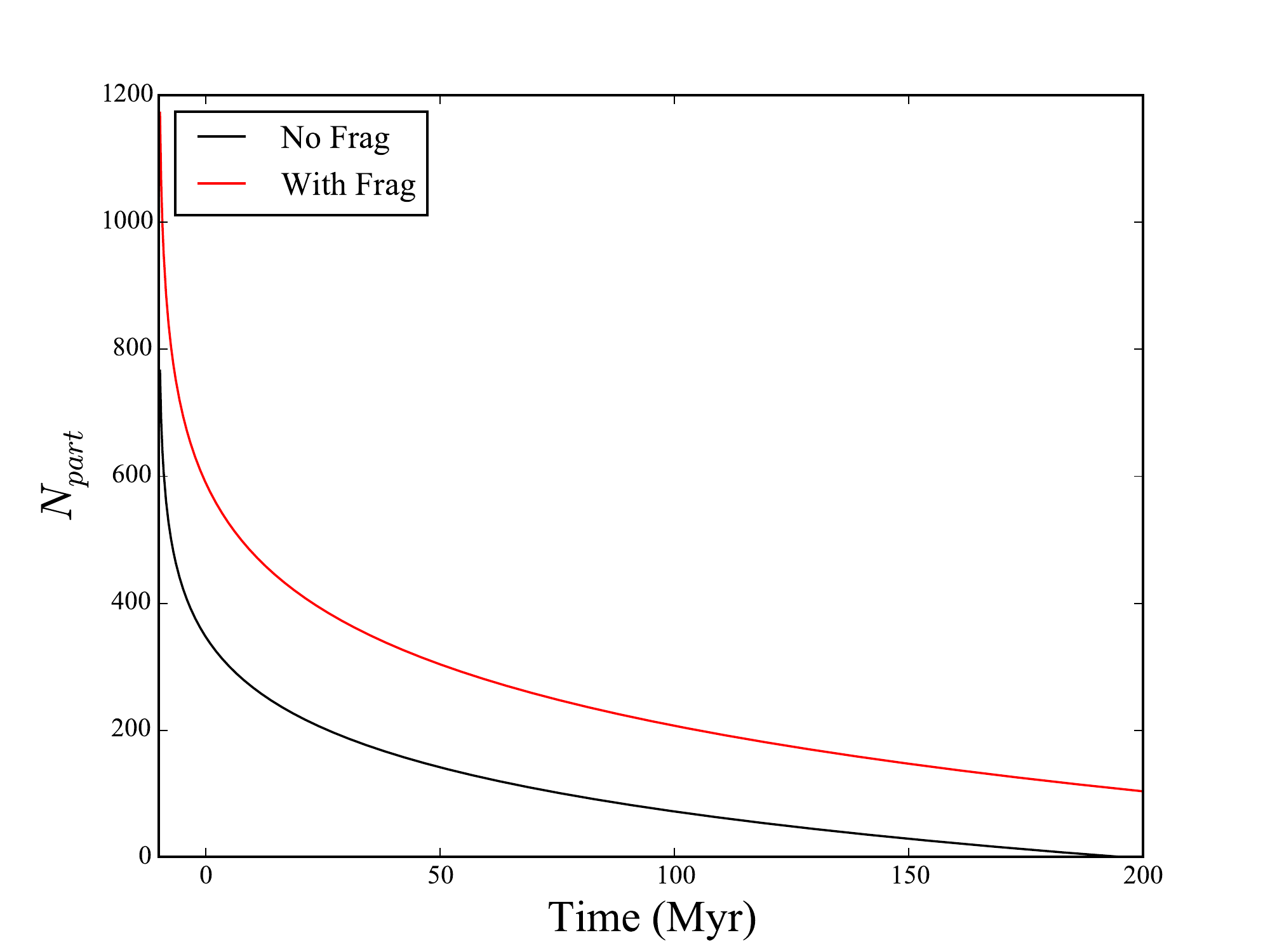}
	\qquad
	\includegraphics[width=.50\textwidth]{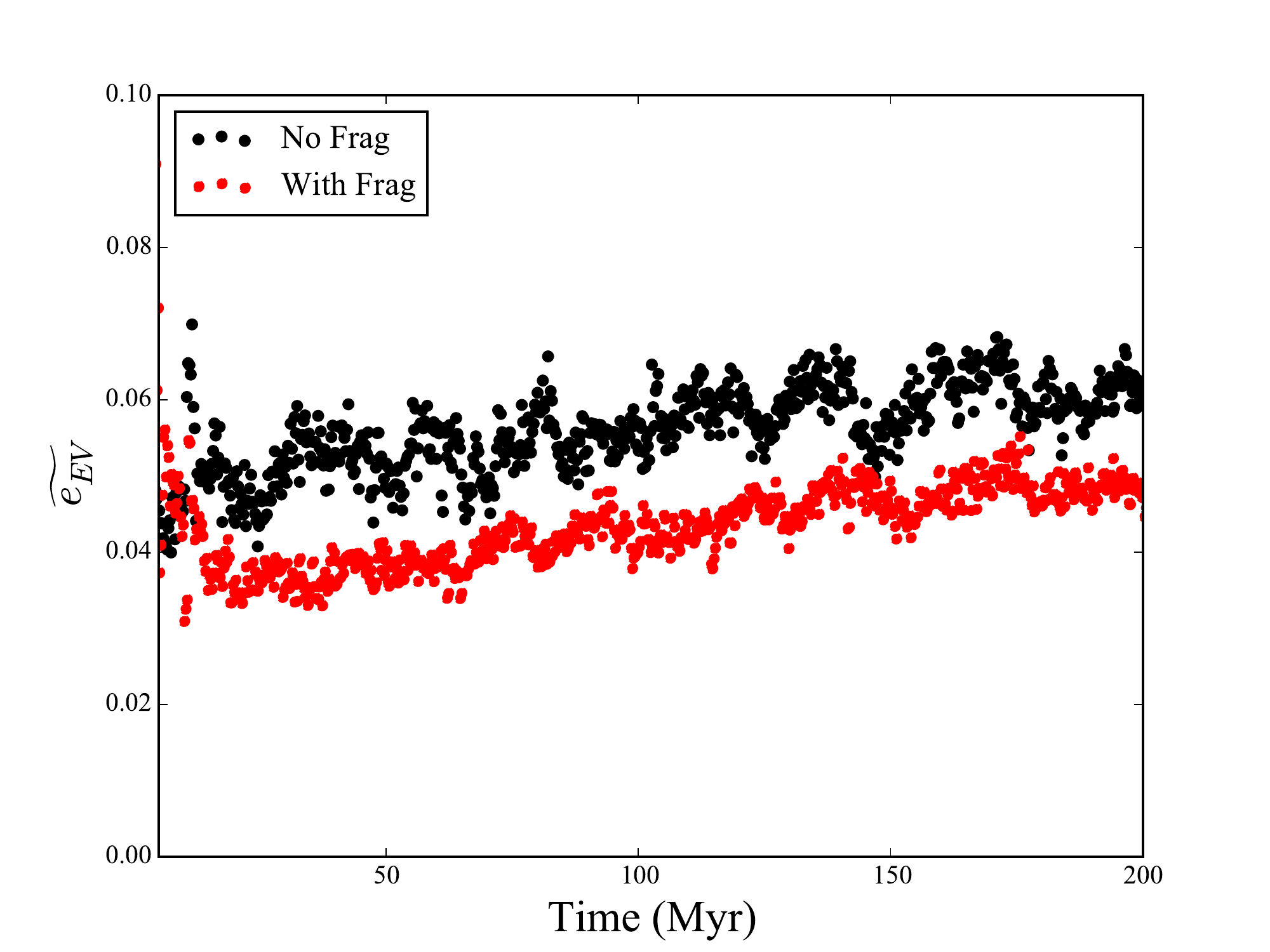}
	\qquad
	\includegraphics[width=.50\textwidth]{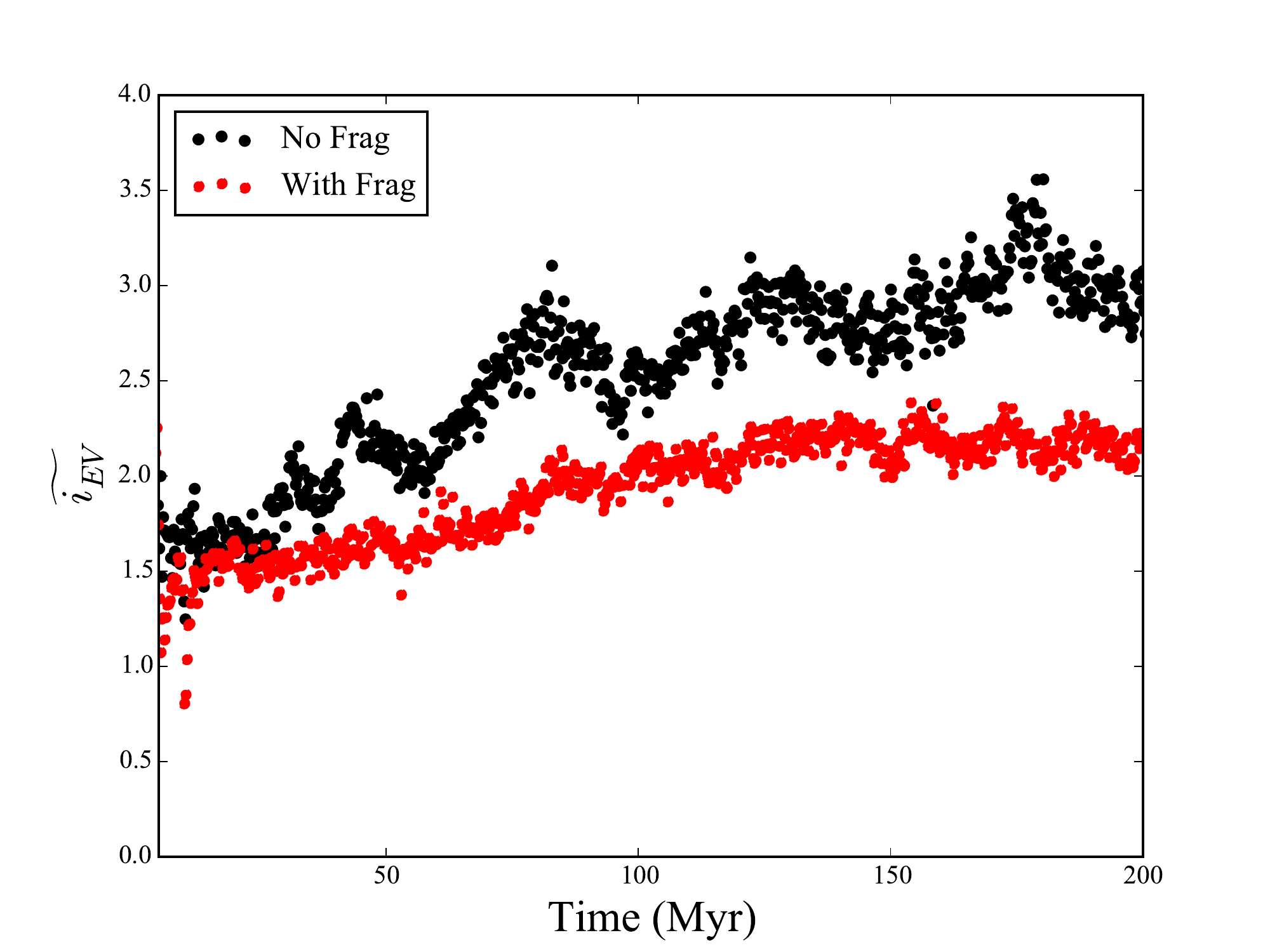}
	\caption{Comparison of disk properties in our instability simulations that include fragmentation (this work) and those performed using a conventional integration scheme (Paper 1).  The red and black lines in the upper panel plot an exponential fit of the average total particle number in the terrestrial forming disk with respect to time.  The red and black points denote mean eccentricities (middle panel) and inclinations (bottom panel) for growing Earth and Venus analogs.}
	\label{fig:e_i}
\end{figure}

In this study, we find that the fragmentation process has much the same effect as an overly-early instability.  The total particle number in the disk stays consistently higher for longer in the fragmentation simulations because of particle addition via fragmenting collisions and accounting for hit-and-run collisions (which would be treated as mergers by conventional integrators).  Though objects that hit-and-run typically go on to merge later in the simulation, the effect is still significant.  In fact, hit-and-runs account for $\sim$30$\%$ of all collision events in our different standard simulation sets, and $\sim$40$\%$ in the various annulus runs.  The net effect is greater dynamical friction in the new fragmentation simulations than in Paper 1.  To demonstrate this, the upper panel of figure \ref{fig:e_i} plots an exponential fit of the average particle number for all instability simulations in this paper versus the corresponding set of simulations from Paper 1 (note, in this plot t$=$0 corresponds to the instability time).  In the bottom two panels, we plot the average eccentricity and inclination of all eventual Earth and Venus analogs (defined in this study as fully formed planets with a$<$1.3 and m$>$0.6 $M_{\oplus}$; see section 2.4.1) at each simulation data output point.  Since embryo orbits are constantly being damped in the disk, it is more difficult for the excited giant planets to remove material from the Mars-forming region than in Paper 1.

A significant percentage ($\sim$35$\%$) of our instability systems form no Mars analog; instead leaving behind a handful of planetesimals or collisional fragments in the region with masses of order 2-5$\%$ the actual mass of Mars.  Though $\sim$30$\%$ of the remaining systems that do form Mars analogs via embryo accretion form planets less massive than Mars (over half are within our success criterion mass of 0.3 $M_{\oplus}$), the instability's tendency to totally inhibit Mars' formation is a potential weakness of the early instability scenario.  We note that this effect is most pronounced in our 1.0 Myr instability delay set.  $\sim$47$\%$ of these systems fail to grow a Mars analog via embryo accretion.  Because the average embryo mass is smaller, collisional grinding in the region is more efficient.  The average impact velocity for fragmenting collisions is 12$\%$ higher in the 1.0 Myr batch than in the 5.0 and 10.0 Myr batches.  Thus more excited fragments are produced and growth is inhibited.

\begin{figure}
\includegraphics[width=.50\textwidth]{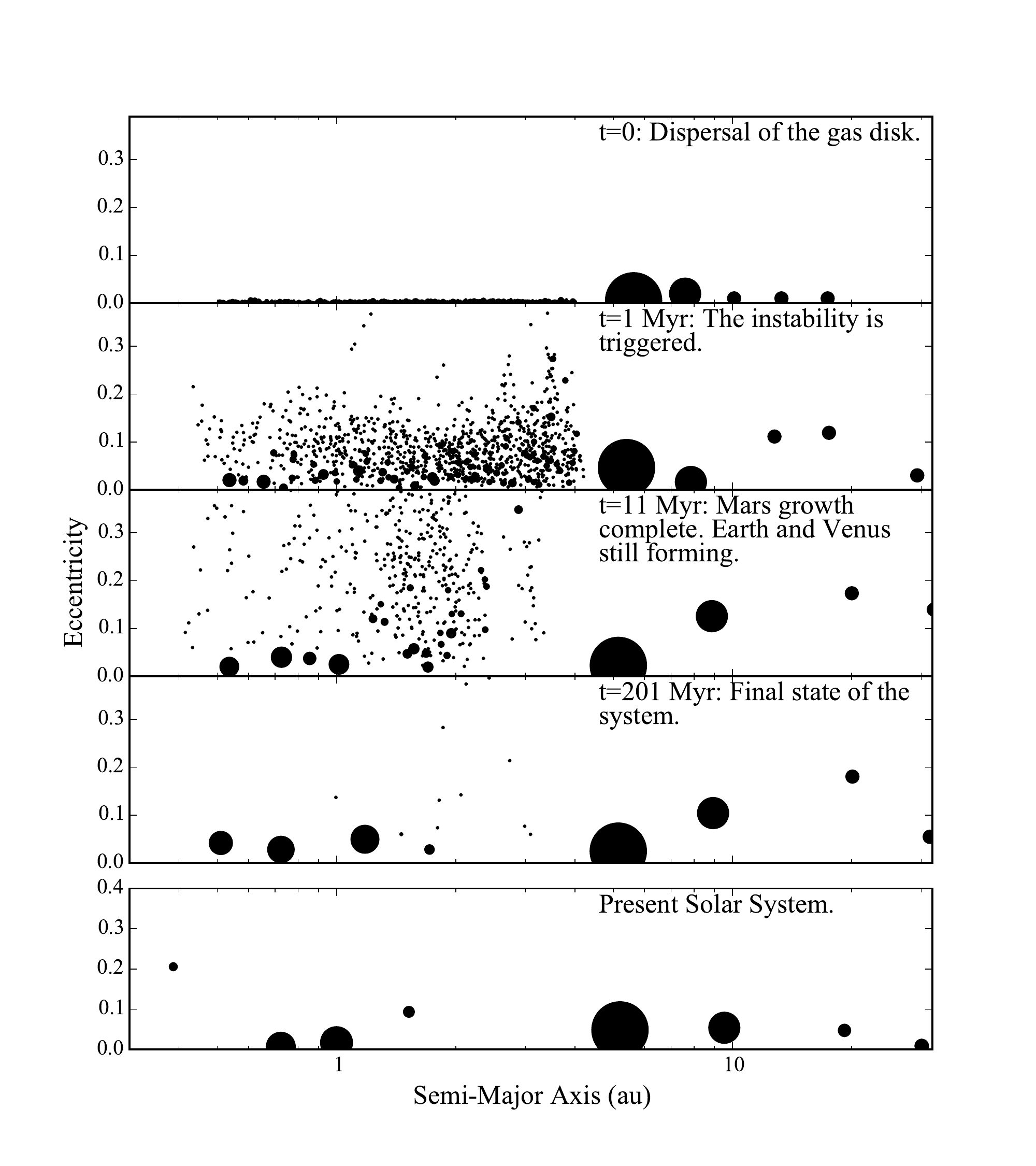}
\caption{Semi-Major Axis/Eccentricity plot depicting the evolution of a successful system in the standard/5GP/1Myr batch.  The size of each point corresponds to the mass of the particle (because Jupiter and Saturn are hundreds of times more massive than the terrestrial planets, we use separate mass scales for the inner and outer planets).  The final terrestrial planet masses are 0.52, 0.68, 0.76 and 0.08 $M_{\oplus}$ respectively.}
\label{fig:time_lapse}
\end{figure}

In general, accounting for collisional fragmentation tends to result in worse Mars analog masses, and more realistic terrestrial eccentricities and inclinations.  The Mars analogs in our fragmentation runs are consistently larger than those formed in Paper 1.  However, they are still dramatically smaller than those in our standard runs without an instability, and well within our established criteria for success (section 2.4.1).  Additionally, as we showed in Paper 1, the majority of our systems with larger Mars analogs experience instabilities that fail to sufficiently excite Jupiter's eccentricity.  In many instances, including the effects of collisional fragmentation results in producing fully evolved systems that are better matches to the actual solar system.  In figure \ref{fig:time_lapse}, we plot an example of the evolution of such a successful system where the final orbits and masses of both the inner and outer planets are most akin to the present solar system.  Though Saturn and Uranus are over-excited, this particular system is highly successful at matching the low orbital eccentricities of the modern terrestrial system.  We discuss this effect in greater detail in section 3.4.

\subsection{Mars' Formation Timescale}

In Paper 1, we argued that an early instability provides a natural explanation for the disparity between the inferred geological accretion timescales of Earth and Mars.  Since the perturbative effect of the excited giant planet orbits (particularity Jupiter and Saturn) is most pronounced in the Mars-forming region and the asteroid belt, Mars' accretion is essentially ``shut off'' by the instability.  In that work, we noted that $\sim$40$\%$ of Mars analogs underwent no impacts with embryos following the instability time.  In this study, we find that effect to be more pronounced due to the integrator's ability to account for hit-and-run collisions.  Depending on the relative velocities and masses involved in such collisions, either all, some or none of the projectile material can be accreted (or re-accreted) over the course of the simulation.  Some objects undergo tens of repeated hit-and-run collisions with one another.  Through this process, it is possible to have a net erosive effect on the larger body.  

Since the parameter space of collisional scenarios is quite complex (number of repeated hit and run collisions, number of initial bodies involved, number of fragments produced, number of fragments involved in subsequent interactions, etc), it is difficult to encapsulate this statistically.  Nevertheless, hit-and-run collisions are extremely common collisional outcomes in our simulations ($\sim$40$\%$ for all collisional interactions in the annulus sets, and $\sim$30$\%$ in the standard sets).  The result is that, for our fragmentation instability systems, $\sim$60$\%$ of Mars analogs undergo no complete accretion events with other embryos following the instability time.  For this reason, our fragmentation runs have much higher rates of satisfying success criterion B (the formation timescale of Mars) than in Paper 1 (see table \ref{table:results}).  Furthermore, Mars analogs in this study accrete at least 90$\%$ of their mass an average of 56 Myr faster than Earth analogs (as compared to 39 Myr in Paper 1 systems).

\subsection{Collisional Evolution of the Asteroid Belt}

As we discuss in section 2.4.3, it is difficult to study the detailed structure of the asteroid belt using 0.025 $M_{\oplus}$ embryos and 0.0025 $M_{\oplus}$ planetesimals.  Even our minimum fragment mass of $\sim$0.0055 $M_{\oplus}$ is 37 times that of the largest asteroid in the belt, Ceres (see discussion in section 2.4.3).  In spite of our coarse asteroid belt resolution, two general trends are obvious.  First, our fragmentation systems have consistently higher rates of meeting success criterion D (leaving behind no embryos in the asteroid belt) than in Paper 1.  Second, our new simulations have lower rates for satisfying criterion E (ratio of main belt asteroids with $a<2.8$ au above to those below the $\nu_{6}$ secular resonance).  We find that this is because collisional fragmentation tends to prevent asteroids from growing larger and accreting in to planets after they are excited by the instability; often times completely shattering them in to many smaller fragments.  Because our fragmentation simulations' embryos stay smaller for longer, they are more easily shattered and destroyed completely in fragmenting collisions.  The main belt's population of relatively young collisional families suggests that it's structure has evolved significantly since the epoch of planet formation \citep{bottke06,walsh13,bottke15,dermott18}.  However, it is extremely difficult to reconstruct the primordial size distribution from the current observed population \citep{bottke05a,bottke05b,delbo17}.  Yet empirical evidence suggests that collisional fragmentation is a dominant process in sculpting the belt's structure (\citet{dermott18} argued that 85$\%$ of all inner main belt asteroids originate from just 5 families).  Since traditional N-body routines used to study planetary dynamics treat all collisions as perfectly accretionary \citep{duncan98,chambers99}, studying these processes numerically has been difficult until recently \citep{chambers13,walsh16}.

Collisions of any kind, including those of the fragmenting variety, are less frequent in the asteroid belt than in the inner solar system due to the lower surface density of material, longer accretion timescales, and higher degree of orbital excitation.  However, when collisions do occur, it is easier for the ejected fragments to survive in the asteroid belt without being re-accreted.  This is because our integrators fragment ejection velocity (set to $\sim$5$\%$ greater than the mutual escape velocity in this study) represents a higher percentage of the mean orbital velocity in the region.  Therefore, it is easier for fragments in the asteroid belt to be ejected on to orbits where they interact less frequently with the original target embryo.  Indeed, over 11$\%$ of all surviving asteroids in our instability systems are collisional fragments.  Additionally, the inclusion of collisional fragmentation further lengthens the accretion timescale in the asteroid belt.  This makes it more difficult for embryos in the asteroid belt to grow in to larger, planet-massed objects; resulting in higher success rates for criterion D.

The small number statistics involved in analyzing each asteroid belt (most contain less than about 30 asteroids) individually can make our calculated $\nu_{6}$ ratios somewhat uncertain.  However, we note that over half of the surviving asteroid belt collisional fragments in our instability simulations are on orbits above the $\nu_{6}$ resonance.  Given the highly excited fragment ejection orbits that are possible in the asteroid belt, this seems to make sense.  However, the vast majority of these asteroids are also on highly eccentric, Mars-crossing orbits.  If we remove fragments on Mars-crossing orbits, only 8$\%$ of the remaining fragment asteroids are above the $\nu_{6}$ resonance.  This is in good agreement with the actual asteroid belt's inclination structure, where the ratio of objects above to those below the $\nu_{6}$ secular resonance is 0.08.  Because collisions occur more frequently at lower inclinations and eccentricities, those asteroids are more likely to be broken up and populate the low inclination parameter space with fragments.  In contrast, over half of all primordial asteroids (embryos and planetesimals) finish the integration above the $\nu_{6}$ resonance.

Though future simulations similar to those in \citet{deienno18} and \citet{clement18_ab} are required to validate the effect of a full early instability on the asteroid belt, our fragmentation simulated asteroid belts are encouraging for three reasons.  First, the additional dynamical friction provided by collisional fragments does not have an appreciable effect on the overall depletion in the Asteroid Belt.  Our new asteroid belts consistently deplete the region at the 90-99$\%$ level.  Second, fragmenting and hit-and-run collisions significantly reduce the rate of planet formation in the asteroid belt.  Finally, non Mars-crossing collisional fragments tend to preferentially populate the inclination parameter space below the $\nu_{6}$ secular resonance.

\subsection{AMDs and RMCs}

\begin{figure}
\includegraphics[width=.50\textwidth]{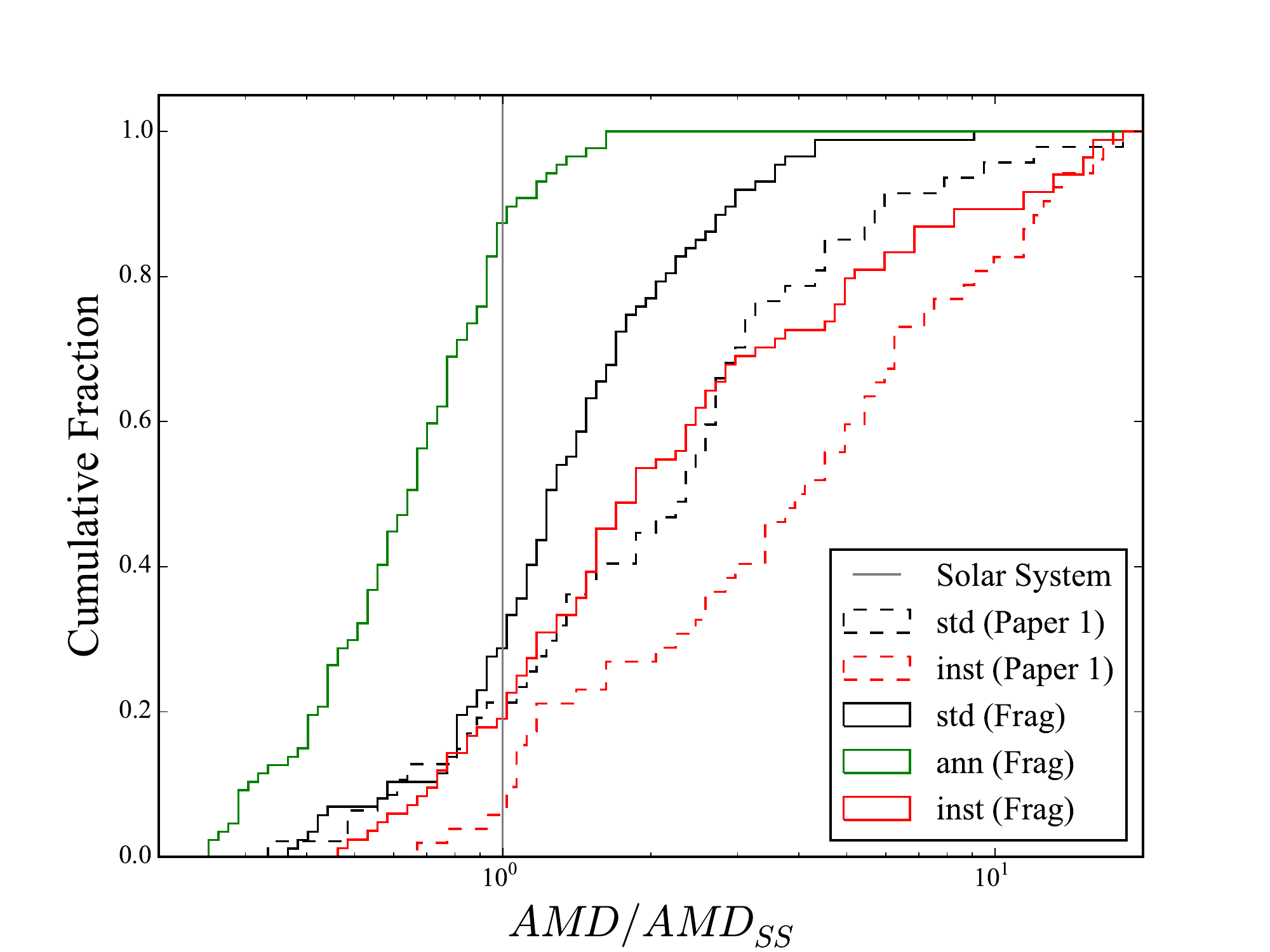}
\qquad
\includegraphics[width=.50\textwidth]{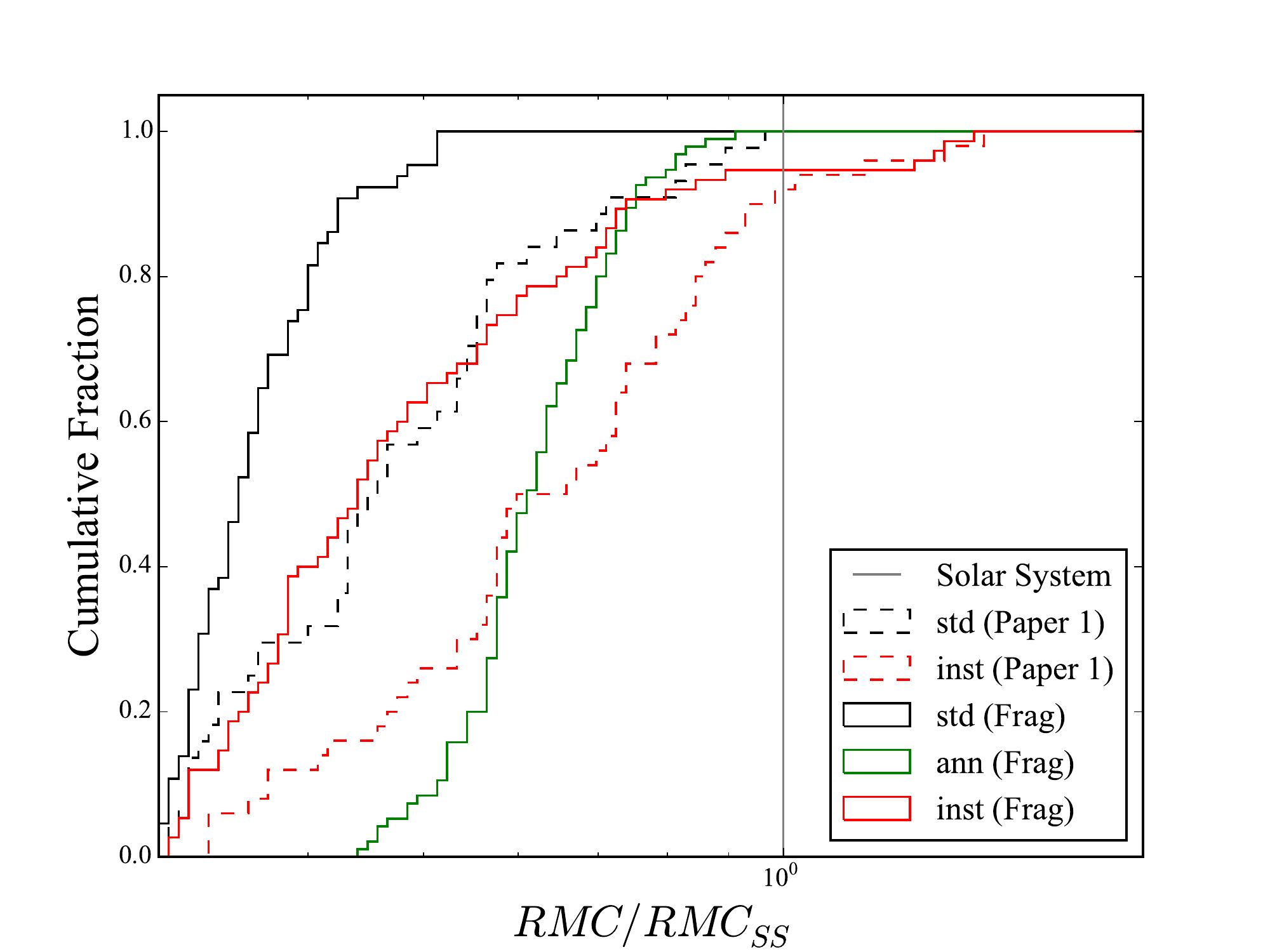}
\caption{Cumulative distribution of normalized AMDs (top panel) and RMCs (bottom panel) for the terrestrial systems formed in our various fragmentation simulation sets (solid lines) compared with our results from Paper 1 (dashed lines). The grey vertical line corresponds to the solar system value (1 by normalization).  Each statistic is calculated utilizing all terrestrial objects that complete the simulation with a$<$2.0 au and m$>$0.055 $M_{\oplus}$, and normalized to the modern solar system value.}
\label{fig:amd_rmc}
\end{figure}

When we scrutinize the systems formed using our fragmentation code, we observe the same general trend of lower final system AMDs as in \citet{chambers13}.  Including fragmentation in our calculation results in a substantial drop in the AMDs of the fully formed terrestrial systems (for a complete discussion consult \citet{chambers13}).  In figure \ref{fig:amd_rmc} (top panel) we plot the cumulative distribution of AMDs for our various simulation sets (standard, annulus and instability) compared with the standard and instability sets from Paper 1.  Though the majority of all the simulation sets besides the annulus runs still possess AMDs larger than the current solar system value, the solar system falls closer to the heart of our new fragmentation AMD distributions.  In Paper 1, fewer than 10$\%$ of our instability systems had AMDs less than that of the modern solar system.  In contrast, over 25$\%$ of the instability systems from our present study have AMDs less than the solar system value.  Since the solar system's terrestrial architecture is well within the spectrum of outcomes that we observe, we argue that accounting for collisional fragmentation is a compelling solution to the terrestrial excitation problem.  Furthermore, given studies that have shown terrestrial AMDs to evolve by a factor of two in either direction over Gyr timescales \citep{laskar97,agnor17}, it may not be necessary for the system AMD to be precisely matched after 200 Myr of planet formation.

In addition to significantly de-exciting the orbits of the fully formed terrestrial planets, accounting for collisional fragmentation also tends to yield systems with better RMCs.  We plot the cumulative distribution of RMCs for our different simulation sets (and Paper 1 systems) in the bottom panel figure \ref{fig:amd_rmc}.  The runs without a Nice Model instability (standard set) from both Paper 1 and our present study consistently yield low RMC values due to the abundance of over-massed Mars analogs and large planets in the asteroid belt.  Our annulus simulations also consistently possess low RMCs because their final architectures often consist of several tightly packed, under-massed planets.  In contrast, more widely spaced systems of larger planets are more common in the standard sets.  These differences reflect the differences in initial mass distribution between the two sets.  

In general, our instability sets consistently yield the best RMC values.  Though the instability sets both with and without fragmentation have similar fractions of systems with RMCs greater than and less than that of the solar system, the median fragmentation system value is closer to the solar system than in Paper 1 (0.73 as opposed to 0.62).  However, many of the fragmentation instability simulation RMCs are lower due to slightly larger Mars analogs (section 3.1).  

In addition to orbits being damped via angular momentum exchange during hit-and-run collisions, dynamical friction between fragments generated in collisions also plays a role in improving system AMDs and RMCs (see discussion in section 3.1 and figure \ref{fig:e_i}).  Though this process tends to result in larger Mars analogs on average, our sample of fully formed systems contains multiple runs that simultaneously meet our constraints for terrestrial system mass distribution and AMD (criterion A and G).  In particular, systems with more realistic Earth/Venus spacings ($\Delta$$a_{EV,SS}=$ 0.28 au) are able to survive 200 Myr of planet formation without combining into a super-Earth.  Indeed, the median semi-major axis spacing between Earth and Venus analogs is 10$\%$ less in our instability fragmentation simulations than in Paper 1.  This is also true for the standard set, where the fragmentation simulations' $\Delta$$a_{EV}$ is 11$\%$ less than in Paper 1 systems.  Thus, accounting for collisional fragmentation leads to forming systems that are better matches to the actual terrestrial system in terms of the Earth/Venus spacing, and total AMD.

\subsection{Annulus Initial Conditions}

Since the annulus simulation set consistently outperforms our other sets in many of our analyses (specifically the masses of Mars analogs, the AMD of the terrestrial system, and our total inner solar system structure success criterion A; see figures \ref{fig:totalplot}, \ref{fig:mars} and \ref{fig:amd_rmc}), the answer to the question of fragmentation's effect on the setup is fairly straightforward.  Our simulations indicate that collisional fragmentation is not a significant impediment to the success of the low mass asteroid belt or Grand Tack models; both of which assume a truncated terrestrial disk similar to our annulus initial conditions.  In fact, collisional fragmentation could provide a mechanism for populating a primordially empty asteroid belt \citep{izidoro15} with silicate-rich material from the inner solar system.  Indeed, 48$\%$ of all final surviving asteroid belt objects in our annulus simulation set are implanted fragments (though $\sim$90$\%$ of them are on highly eccentric, Mars crossing orbits).  

In Paper 1 we showed that the Nice Model instability is efficient at disturbing and depleting the terrestrial disk mass in the Mars-forming region and beyond ($a>1.3$ au).   Since our annulus simulation set begins with all the terrestrial forming mass concentrated in a narrow annulus with $0.7<a<1.0$ au, we do not expect the instability to have a substantial effect on the planet formation process in these simulations.  Indeed, the annulus/XGP/10Myr simulations have similar success rates for criteria A and A1 (the bulk architecture of the inner solar system) as the annulus simulations (which do not include any giant planet evolution).  The instability set is, however, less successful at satisfying the terrestrial AMD requirement (criterion G).  This is expected given the tendency for the terrestrial planets to be excited by the evolving giant planet orbits in simulations of the Nice Model \citep{bras09,agnorlin12,bras13,kaibcham16}.  Furthermore, since the annulus set begins with a larger number of embryos on Hill-crossing orbits, they evolve much quicker (see discussion below), and are at a more advanced stage of evolution when the instability ensues.  Therefore, the effect of fragmentation and hit-and-run collisions on AMDs (section 3.4) is less drastic in the annulus instability set.

\begin{figure}
\includegraphics[width=.50\textwidth]{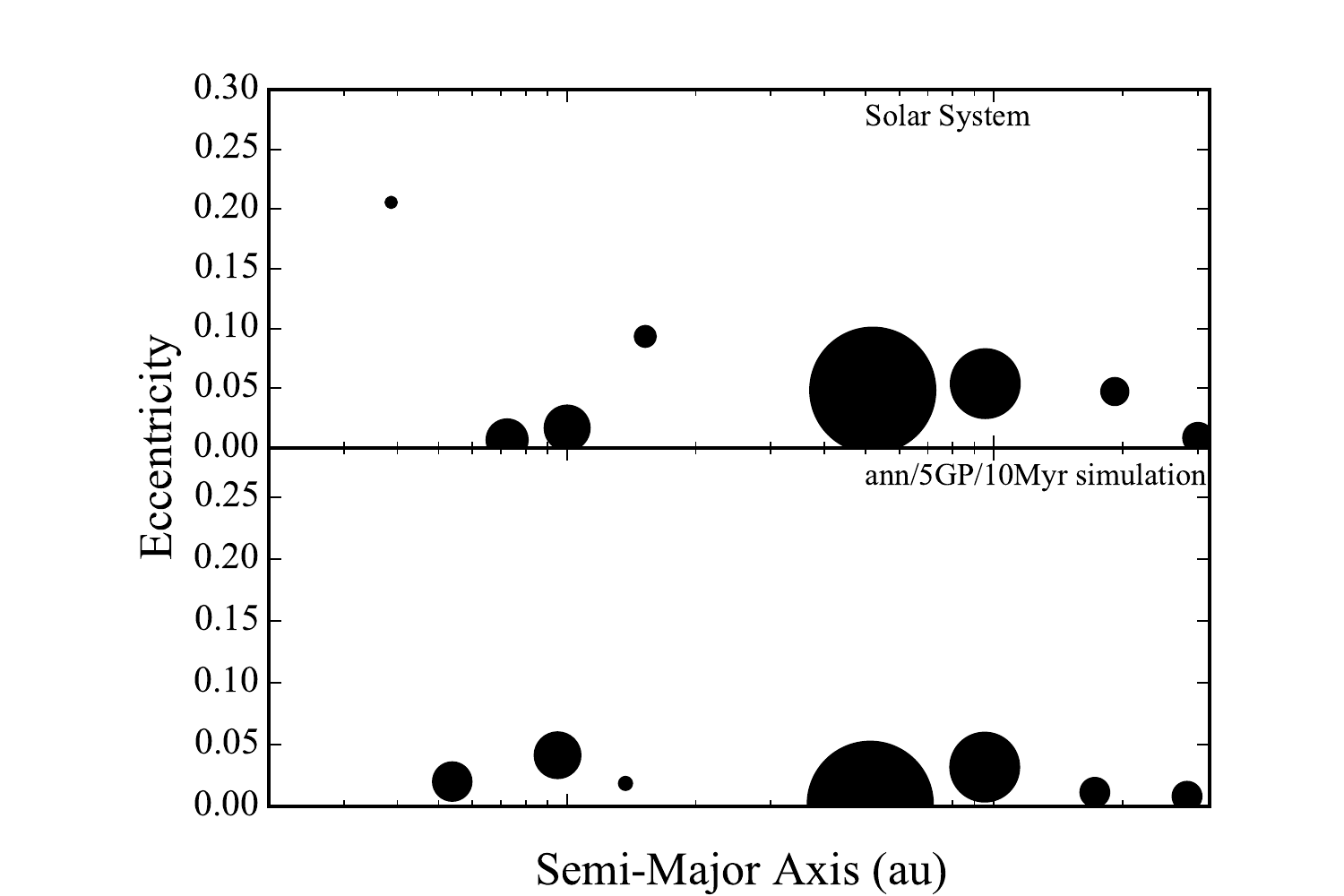}
\caption{Semi-Major Axis/Eccentricity plot depicting a successful system in the annulus/5GP/10Myr batch (bottom panel), compared with the actual solar system (top panel).  The size of each point corresponds to the mass of the particle (because Jupiter and Saturn are hundreds of times more massive than the terrestrial planets, we use separate mass scales for the inner and outer planets).  The final terrestrial planet masses are 0.73, 1.04 and 0.076 $M_{\oplus}$ respectively.}
\label{fig:gt}
\end{figure}

Overall, the results of the annulus/5GP/10Myr and annulus/6GP/10Myr simulations do not differ substantially from the annulus set.  These sets frequently produce systems that broadly match the observed characteristics of the real inner solar system.  An example of a successful simulation is plotted in figure \ref{fig:gt}.  These results should be taken in the appropriate context given the over-simplification of our annulus initial conditions.  The important conclusion is that timing the orbital instability in conjunction with terrestrial planet formation (regardless of the terrestrial disk initial conditions or prior evolutionary scheme) is a viable solution to the problem of terrestrial system disruption in a late instability scenario (particularly the high rates of collisions and ejections).

Finally, we note that the Earth analogs ($m>0.6$ $M_{\oplus}$, $0.85<a<1.3$ au) in all of our annulus simulation sets have much lower rates of satisfying criterion C (geological formation timescale of Earth; only 27$\%$ of annulus systems).  Mars analogs in the annulus set form an average of just 11 Myr faster than their counterpart Earth analogs (as compared to 56 Myr in our instability sets).  Given the higher initial surface density, faster accretion timescales are expected in this set.  Indeed, 3 of our annulus Earth analogs accrete 90$\%$ of their mass in less than 20 Myr.  To test whether this is a consequence of our use of 400 self-gravitating embryos, and no planetesimals (see also \citet{jacobson14}), we perform an additional batch of 25 simulations that include planetesimals.  The setup for these integration is identical to the annulus set's initial conditions, with the exception of the disk mass being divided in to 1 $M_{\oplus}$ of 50 equal-mass embryos and 1 $M_{\oplus}$ of 500 equal-mass planetesimals.  Indeed, the lack of a bi-modal initial disk mass distribution seems to be the cause of the rapid growth of Earth analogs in our annulus set.  83$\%$ of Earth analogs in our new, embryo/planetesimal annulus set meet success criterion C.  Furthermore, Earth analogs take 36.2 Myr longer to form than corresponding Mars analogs in this new set.

\section{Future Work}

We have shown that the early instability scenario \citep{clement18} and the various annulus models (Grand Tack and low mass asteroid belt; \citet{walsh11,ray17sci}) are all viable when scrutinized against multiple standard success criteria (table \ref{table:crit}).  In the subsequent sections, we offer a brief synopsis of how future studies can use different constraints and higher resolution simulations to accurately distinguish between models.

\subsection{Late Terrestrial Bombardment}

The robust set of samples returned by the Apollo missions yielded ages for several large impact basins between $\sim$ 4.3 and 3.7 Gyr \citep{heiken74,fritz14,zellner17}.  Furthermore, the low maturity index of regolith overlying orange ash at Shorty Crater has been interpreted to imply that the regolith could not have existed in the presence of a significant flux of micro-meteor impacts \citep{heiken74,morris78,schmitt14}.  Since these ash samples returned ages around 3.5 Gyr, the micro-meteor flux on the Moon must have been essentially zero when they were formed.  The ages of  impact pulverized lava flow returned in lunar core sample tubes is also consistent with the bombardment being totally complete about 3.5 Gyr ago \citep{schmitt14}.  Thus, regardless of the early impact distribution (smooth decline or LHB), the majority of the larger basins must form before $\sim$3.7 Gyr ago, and the inner solar system must be essentially clear of debris before $\sim$3.5 Gyr ago \citep{fritz14,zellner17}. However, it should be noted that the topic of the Moon's geological history is extremely complex, with many competing models and contradicting pieces of data.  In fact, evidence exists that the micro-meteor flux has been variable throughout the last 3.5 Gyr \citep{johnson16}, and has been relatively high for about the last 75 Myr \citep{schmitt17}.  \citet{nesvorny10} showed that the majority of this modern flux is fueled from Jupiter family comets.  It is thus difficult to correlate ancient micro-meteor activity with debris clearing in the inner solar system.  As for the formation of the larger basins, collisional grinding has been shown to quickly break up large enough projectiles \citep{bottke07_lhb}.

\subsection{Extended Simulations}

To compare our simulations with the record of lunar samples, we must express time relative to CAI formation (calcium aluminum-rich inclusion; approximately time zero in our simulations as discussed in Paper 1).  Thus the 3.7 Gyr basin ages corresponds with $\sim$800 Myr after CAI, and the 3.5 Gyr zeroing of the micro-meteor flux converts to $\sim$1 Gyr after CAI.  This provides an interesting constraint for our terrestrial evolution simulations.  Indeed, our fragmentation simulations typically finish with a population of small, unstable debris in the inner solar system after 200 Myr of integration (\citet{chambers13} noted a similar lengthening of complete accretion timescales in the inner solar system).  The instability simulations in Paper 1 had an average of 1.1 additional objects smaller than Mercury in the inner solar system ($a<2.0$ au), as compared to 2.9 in our new fragmentation simulations.  This gives average leftover masses consistent with the the $\sim$0.05 $M_{\oplus}$ necessary to produce the highly siderophile element (HSE) signature in the Earth's mantle (\citet{raymond13}; see also \citet{brasser16_lhb}).

\begin{figure}
	\centering
	\includegraphics[width=.50\textwidth]{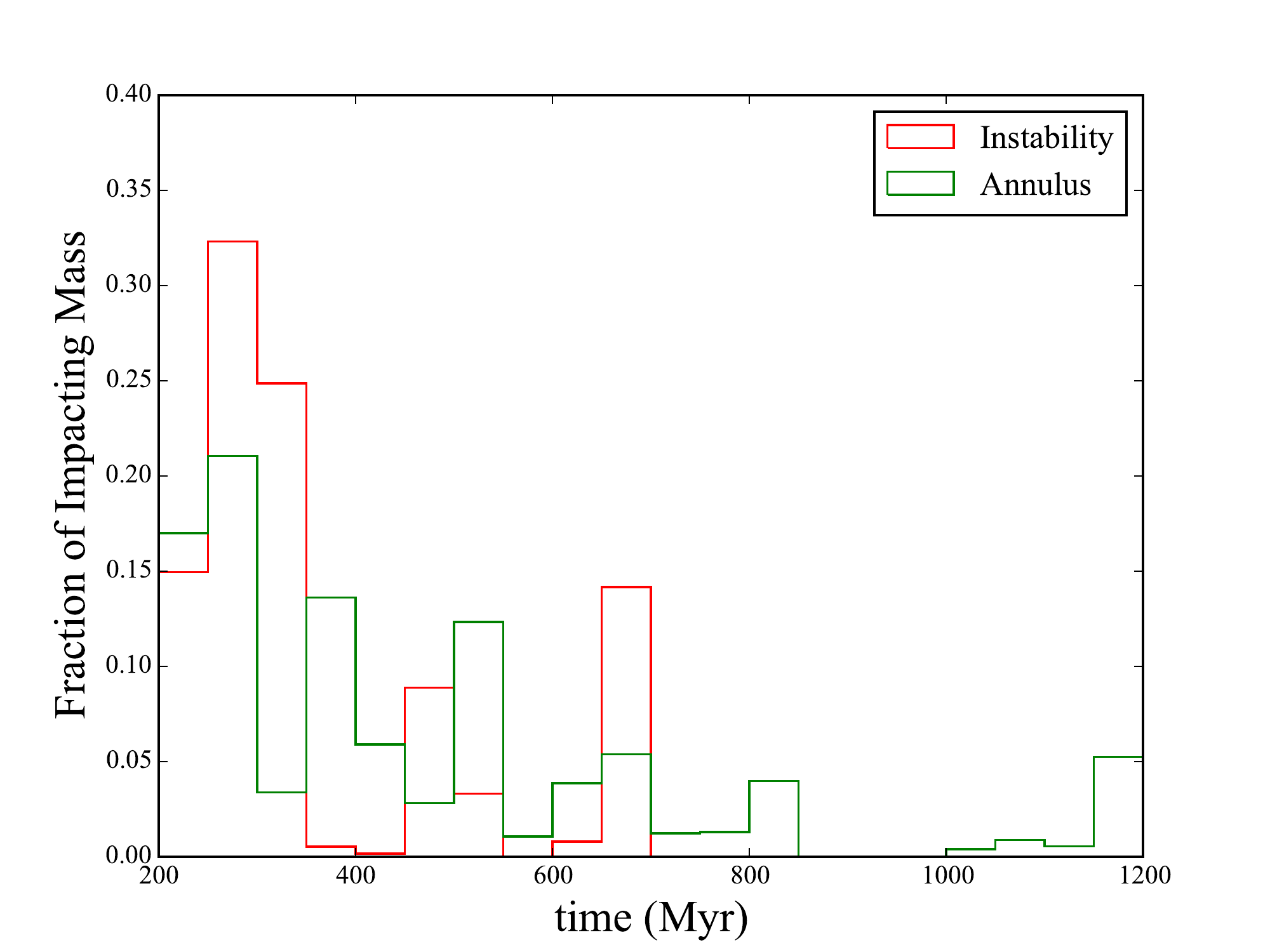}
	\caption{Relative late impact rates with respect to time on Venus, Earth and Mars analogs in our 1 Gyr extended annulus and instability integrations.}
	\label{fig:lhb}
\end{figure}

To investigate the late bombardment in these systems, we randomly select 32 systems with higher than the median number of debris particles from each of our most successful simulation sets (our standard instability set and our annulus set without giant planet evolution), and integrate them for an additional 1 Gyr.  These extended simulations utilize the same integrator and time-step as the original terrestrial planet formation runs.  To limit the computational cost of these integrations, we remove all ice giants, Kuiper Belt objects, and asteroids with perihelion greater than 2.0 au.  Objects with masses less than 0.055 $M_{\oplus}$ do not interact gravitationally with one another.  In figure \ref{fig:lhb}, we plot the relative late mass delivery rate on all Venus, Earth and Mars analogs for each simulation set.

\subsection{Late Impacts}

The particles in our simulations are too massive to adequately study the process of late terrestrial bombardment.  The lower range of our debris particles' mass distribution (ranging from $\sim$5$\%$ the mass of the Moon to several lunar masses) is two orders of magnitude larger that that of the objects that formed the lunar basins.  The mass of the impactor that formed the Imbrium basin, for example, is estimated to have been just $\sim$0.03$\%$ that of the Moon \citep{schultz16}.  However, the clearing rates of our instability set (no late impacts after 700 Myr) is noticeably different than the annulus set (the distributions were determined to be significantly different via a Kolmogorov-Smirnov test yielding a p-value of 2.23x$10^{-5}$).  This is because the debris particles in the annulus set are not as dynamically excited ($\bar{e}_{ann}=$0.24, $\bar{i}_{ann}=$10.7$^{\circ}$) as the instability set's debris ($\bar{e}_{instb}=$0.31, $\bar{i}_{instb}=$21.3$^{\circ}$).  Thus the debris objects in the annulus set are more likely to be on less eccentric, quasi-stable orbits that can survive for nearly 1 Gyr without experiencing an instability.  Indeed, our extended instability simulations lose 36 times as many objects via mergers with the Sun than in the annulus simulations (part of this is due to the giant planets' orbits not being as excited in the annulus set as in the instability set).  

\subsection{HSE Constraints}

We can also scrutinize our extended simulation set against constraints for the late veneer and Earth's HSE inventory \citep{raymond13,jacobson_nat14,morb18}.  Earth analogs (m$>$0.6 $M_{\oplus}$, 0.85$<$a$<$1.3 au) in our instability set accrete an average of 0.006 $M_{\oplus}$ in chondritic (non-fragment) material after the last giant impact as opposed to 0.009 $M_{\oplus}$ in our annus set.  Given the low particle resolution, both of these values are roughly consistent with the 0.003-0.007 $M_{\oplus}$ required to match the modern HSE concentration \citep{walker09}.  However, our instability set Earth analogs accrete two full orders of magnitude more silicate-rich fragment material than the annulus set (0.02 vs 0.0003 $M_{\oplus}$).  Given the small number of simulations and impacts, it is difficult to draw significant conclusions from these values.  Nevertheless, comparing late accretion histories in high-resolution N-body simulations including fragmentation could be a promising means of differentiating between formation models in the future.

\section{Conclusions}

In this paper, we presented the largest ever sample of simulations of terrestrial planet formation in the solar system using an integrator that considers the effects of collisional fragmentation.  In particular, we performed a detailed reinvestigation of the early instability scenario proposed in \citet{clement18} (Paper 1).  Our new simulations of terrestrial planet formation occurring in conjunction with the Nice Model instability consistently outperformed those with no giant planet evolution when measured against a wide range of success criteria.  

Including the effects of collisional fragmentation yielded systems of terrestrial planets with more realistic radial mass distribution profiles (particularly the Earth/Venus spacing), on orbits that were better matches to the solar system in terms of their lower eccentricities and inclinations.  While dynamical friction from the additional collision-generated fragments and angular momentum transfer in hit-and-run collisions tends to damp the orbits of the growing planets, these processes also occasionally inhibit mass-loss in the Mars region during the instability.  As a result, the Mars analogs in our fragmentation simulations of the early instability scenario are somewhat larger than those reported in Paper 1.  This problem is lessened somewhat by moving the instability earlier ($\sim$1.0 Myr after gas dissipation), but such a timing also boosts the probability of entirely preventing Mars' formation.  Still, many of our new instability systems form planets in the Mars region smaller than Mars.  

Because hit-and-run collisions prevent the late growth of Mars analogs, and the constant resupply of fragments lengthens the accretion timescales of Earth analogs, our fragmentation simulations provide better matches to the inferred geological growth histories of the two planets \citep{kleine09,Dauphas11} than in Paper 1.  Finally, we find that collisional fragmentation and hit-and-run collisions play a dominant role in preventing planet formation in the primordial asteroid belt.

The early instability scenario's explanation for Mars' small mass has the advantage of simplicity,  while still relying on the giant planets' influence to help solve the small Mars problem.  The model is consistent with the diminishing evidence for a LHB \citep{zellner17,morb18}, does not involve a dynamically fine tuned delayed instability and, most importantly, saves the terrestrial planets.  Certain geochemical and dynamical constraints including isotopic data from comet 67P \citep{marty17} and constraints on forming the cold classical Kuiper Belt \citep{nesvorny15a} are somewhat at odds with our preferred instability timing of $\sim$1.0 Myr after gas disk dispersal (see full discussion in Paper 1).  However, we argue that our model's ability to preserve the terrestrial system during the giant planet instability is more important than simultaneously reconciling all timing constraints; each of which have their own uncertainties and model dependencies.

Despite the advantages of the early instability model, the processes involved in terrestrial planet formation are highly chaotic and stochastic.  Since observational constraints are limited, it is difficult to rule out one formation model in favor of another.  Furthermore, the various explanations for Mars' small mass (eg: early instability, low-mass asteroid belt, pebble accretion, Grand Tack hypothesis, etc.) could have actually sculpted Mars' early evolution in tandem with one another.  To address this, we performed an additional set of simplified simulations to study the compatibility of the early instability scenario with the truncated disk initial conditions supposed by the Grand Tack and low mass asteroid belt models (a narrow annulus of terrestrial forming material between $\sim$0.7 and 1.0 au; \citet{hansen09}).  These simulations indicated that the annulus setup is compatible with an early giant planet instability.  Additionally, collisional fragmentation does not seem to be a barrier to the success of the initial conditions.

Future work on these topics is required to clarify the advantages and disadvantages of the various proposed terrestrial evolutionary schemes.  In particular, fragmentation-style simulations like those presented in this paper should be analyzed in more detail to see if the cratering record of the Moon is consistent with the tail end of terrestrial planet formation.  This next step might provide a concrete test with which to differentiate between models.

\section*{Acknowledgments}

This material is based upon research supported by the Chateaubriand Fellowship of the Office for Science and Technology of the Embassy of France in the United States.  M.S.C. and N.A.K. thank the National Science Foundation for support under award AST-1615975.  S.N.R. thanks the Agence Nationale pour la Recherche for support via grant ANR-13-BS05-0003-002 (grant MOJO) and acknowledges NASA Astrobiology Institute’s Virtual Planetary Laboratory Lead Team, funded via the NASA Astrobiology Institute under solicitation NNH12ZDA002C and cooperative agreement no. NNA13AA93A.  This research is part of the Blue Waters sustained-petascale computing project, which is supported by the National Science Foundation (awards OCI-0725070 and ACI-1238993) and the state of Illinois. Blue Waters is a joint effort of the University of Illinois at Urbana-Champaign and its National Center for Supercomputing Applications \citep{bw1,bw2}.  Further computing for this project was performed at the OU Supercomputing Center for Education and Research (OSCER) at the University of Oklahoma (OU).  Additional analysis and simulations were done using resources provided by the Open Science Grid \citep{osg1,osg2}, which is supported by the National Science Foundation award 1148698, and the U.S. Department of Energy's Office of Science.  Several sets of simulations were managed on the Nielsen Hall Network using the HTCondor software package: https://research.cs.wisc.edu/htcondor/.

\bibliographystyle{apj}
\newcommand{\sci}{$Science$ }
\bibliography{frag}
\end{document}